\documentclass{aa}  

\usepackage{graphicx}
\usepackage{txfonts}
\usepackage{lscape}
\usepackage{xcolor}     

\begin{document}

   \title{A fast-filament eruption observed in the H$\alpha$ spectral line} 

   \subtitle{I. Imaging spectroscopy diagnostic}

   \author{Denis P. Cabezas\inst{1,2}\fnmsep\thanks{ Corresponding author.}, 
   		Kiyoshi Ichimoto\inst{2}, Ayumi Asai\inst{2}, Satoru UeNo\inst{2}, Satoshi Morita\inst{3}, 
		Ken-ichi Otsuji\inst{3}, \and Kazunari Shibata\inst{2,4}}
   \institute{Institute for Space-Earth Environmental Research, Nagoya University. Furo-cho, Chikusa-ku,
                Nagoya, Aichi, 464-8601 Japan \\
              	\email{denis.cabezas@isee.nagoya-u.ac.jp} \and
	   	Astronomical Observatory, Kyoto University, Kitashirakawa-Oiwake-Cho, Sakyo-Ku,
		Kyoto, 606-8502 Japan \and
		National Institute of Information and Communications Technology, Nukui-Kitamachi, Koganei,
		Tokyo, 181-8795 Japan \and
		School of Science and Engineering, Doshisha University, Kyotonabe, Kyoto, 610-0394 Japan}

   \date{Accepted: \today }

 
  \abstract
   {Solar filament eruptions usually appear to occur in association with the sudden explosive release of
   magnetic energy accumulated in long-lived arched magnetic structures.
   It is the released energy that occasionally drives fast-filament eruptions that can be source regions
   of coronal mass ejections. A quantitative analysis of high-speed filament eruptions is thus essential
   to help elucidate the formation and early acceleration of coronal mass ejections.
   }
   {The goal of this paper is to investigate the dynamic processes of a fast-filament eruption by using
   unprecedented high-resolution full-disk H$\alpha$ imaging spectroscopy observations.
   }
   {The whole process of the eruption was captured in a wide spectral window of the H$\alpha$ line
   ($\pm9.0$~{\AA}), which allowed the detection of highly Doppler-shifted plasma. Applying the
   ``cloud model'' and obtaining two-dimensional optical thickness spectra we derive the Doppler velocity,
   the true eruption profiles (height, velocity, and acceleration), and the trajectory of the filament
   eruption in 3D space.
   }
   {The Doppler velocity maps show that the filament was predominantly blue-shifted. During the main and
   final process of the eruption, strongly blue-shifted materials are manifested traveling with velocities
   exceeding $250~{\rm km~s^{-1}}$. The spectral analysis further revealed that the erupting filament is
   made of multiple components, some of which were Doppler-shifted approximately to $-300~{\rm km~s^{-1}}$.
   It is found that the filament eruption attains a maximum true velocity and acceleration of about
   $600~{\rm km~s^{-1}}$ and $2.5~{\rm km~s^{-2}}$, respectively, and its propagation direction
   deviates from the radial direction.
   On the other hand, downflows manifested as red-shifted plasma close to the footpoints of the erupting
   filament move with velocities $45-125~{\rm km~s^{-1}}$. We interpret these red-shifted signatures as
   draining material, and therefore mass loss of the filament that has implications for the dynamic and
   the acceleration process of the eruption.
   Furthermore, we have estimated the total mass of the H$\alpha$ filament resulting in
   $\sim$$5.4\times10^{15}{\rm g}$.
   }
   {}

   \titlerunning{A fast-filament eruption: I. Imaging spectroscopy}
   \authorrunning{Denis P. Cabezas et al.}

   \maketitle
%

\section{Introduction}
Solar filaments which are seen as dark elongated structures against the bright solar disk or prominences when projected above the
solar limb as bright standing-up strands, are dense and cool ``magnetic clouds'' immersed in the chromosphere and hot corona. They
are best observed in cool spectral lines such as hydrogen and helium that originate at chromospheric temperatures. This indicates
that filaments are made of relatively low temperature ($\sim$$10^{4}~{\rm K}$) and high density ($\sim$$10^{11}~{\rm cm^{-3}}$)
plasma, that is, about two orders of magnitude cooler and denser than the surrounding tenuous coronal environment. Depending on
the morphology, dynamic properties, and their relative location on the solar disk, solar filaments are generally classified into
three categories: quiescent, intermediate, and active region filaments \citep[][and references therein]{mackay2010, engvold2015}.

Because quiescent filaments form along the magnetic polarity inversion line and are confined only in some specific regions of the
solar atmosphere \cite[]{martin1994,tandberg-hanssen1995}, it is believed that they are sustained in equilibrium against gravity
by the horizontal component of the ambient magnetic field and are thermally shielded from the encompassing hot corona
\citep{tandberg-hanssen1970, tandberg-hanssen1995, engvold2015, hillier2018}. This means that any change or reconfiguration of the
magnetic field harboring a filament may lead to the loss of equilibrium and the eventual eruption of the filament.
On the other hand, mass unloading or mass draining has also been pointed out as an important factor that could induce the loss of
equilibrium and subsequent eruption of filaments. Observational evidence of mass unloading before a filament eruption was
presented in early works \citep[e.g.][]{seaton2011, huang2014, li2017, jenkins2018, doyle2019}, and quite recently analytical and
numerical studies were also conducted \citep[e.g.][]{vrsnak2019, tsap2019, fan2020}. Accordingly, it is reasonable to believe that
mass drainage that could appear during filament and prominence eruptions may also exert influences on the acceleration process.

In many cases observations have revealed that during the pre-eruption stage, filaments exhibit signs of activation followed by a
slow rise and a gentle expansion, thereafter, at a certain point when the filament enters an eruption stage it experiences a
drastic transition from slow to fast expansion \citep[e.g.][]{kahler1988}. Particularly, in the latter stage the system becomes
very dynamic where a large amount of the filament material is strongly accelerated upward to velocities over $100~{\rm km~s^{-1}}$.
Filament eruptions can be accompanied by solar flares and coronal mass ejections (CME), but interestingly most of the erupting
filaments are directly associated with the occurrence of CMEs
\citep[e.g.][and references therein]{gilbert2000,forbes2006, gopalswamy2015}. Therefore, understanding the initiation and
acceleration processes of filament and prominence eruptions can provide reliable clues on the formation, development, and how the
CMEs are launched. In this respect, we can mention new results about eruptive events and CMEs reported by \cite{russano2024} based
on observations of the Metis coronagraph \citep{antonucci2020} onboard the Solar Orbiter \citep{muller2020}.

Several works have dealt with the problem of acceleration of filament eruptions trying to elucidate qualitatively the governing
physical processes involved in the eruptions \cite[e.g.][and references therein]{gilbert2000,vrsnak2001,gallagher2003,torok2005,
williams2005, schrijver2008, cheng2020}. In these works the authors used different fitting models, such as linear, non-linear, or
composite models to replicate the eruption profiles inferred from observations. Although in most case studies the main
acceleration phase could be satisfactorily characterized with exponential forms, it has been found that polynomial and power-law
functions can also yield confidence results \citep{priest2002, schrijver2008}. It is thus clear that different driving processes
are at work during the eruptions, which also means that the mechanism responsible for filament eruptions may vary from event to
event. We leave the theoretical aspects of filament eruptions and results of numerical modeling to Cabezas et al. (2024)
(in preparation, hereafter Paper II).

Quantifying the real magnitude of filaments and prominence eruptions requires an accurate determination of the actual propagation
speed. However, at present most of the studies still make use of the projected motion in the plane-of-sky as the main parameter,
and in some cases assumptions are imposed to estimate the true motion of the erupting material paying no attention to the
line-of-sight component (i.e. directed towards the Earth). Therefore, caution should be taken because quite often it is used only
the plane-of-sky component to draw conclusions, and also adopted in the majority of models to predict the propagation of CMEs.
Although, we should make a distinction between eruptions occurring on-disk and off-limb (or at the solar limb). If an off-limb
eruption has a strong radial velocity component, then its actual propagation speed could be constrained mainly by its apparent
motion in the plane-of-sky. However, the situation of on-disk eruptions requires appropriate treatments because they can be
strongly affected by the line-of-sight component.

The difficulties in deriving the true motion of erupting filaments and prominences rely mainly on the limited spectral window of
conventional instruments, the small field of view in the case of large-aperture telescopes, and the scanning time across the
full-disk image in the case of spectrographs. These limitations prevent the detection of large Doppler-shifted plasma and to track
the whole evolution of the eruption. Here it is worth mentioning the work by \cite{li2005}, who based on H$\alpha$ spectroscopic
observations reported an extreme blue-shift of a filament eruption exceeding $480~{\rm km~s^{-1}}$, and the calculated true
velocity of the eruption was $\sim$$700~{\rm km~s^{-1}}$. Similarly, \cite{penn2000} presented observations of a filament eruption
and showed that the He~{\sc I} line (10830~\AA) was strongly blue-shifted to velocities $200-300~{\rm km~s^{-1}}$.
Recently, \cite{liu2015} and \cite{zhang2019} also found large Doppler-shifted prominence eruptions in spectroscopic observations.
On the other hand, some efforts have been made to recover the true velocity and 3-dimensional trajectory of filament and prominence
eruptions employing spectroscopic observations \citep{zapior2010, zapior2012, schmieder2017, sakaue2018} and multiwavelength
imaging observations \citep{morimoto2003, chae2006, cabezas2017, gutierrez2021}.
Alternatively, there are few reports in which by combining observations from different perspectives (stereoscopy) the velocity
components of filament and prominence eruptions could be estimated \citep[e.g.][]{bemporad2009, li2010, bemporad2011, rees2020,
janvier2023, russano2024}, although this kind of observations are still very sporadic.

In this paper, we report on a remarkable fast-filament eruption detected in a broad range of the H$\alpha$ spectral profile by the
Solar Dynamics Doppler Imager \citep[SDDI;][]{ichimoto2017} onboard the Solar Magnetic Activity Research Telescope
\citep[SMART;][]{ueno2004}. This novel observing system that provides high-resolution full-disk filtergrams in a wide spectral
window of the H$\alpha$ line is particularly suitable for performing imaging spectroscopy analysis of filament and prominence
eruptions that may emerge from any location of the solar disk. Here we focus on the different aspects of the filament eruption
with particular emphasis on the detection of highly Doppler-shifted plasma, the time development of the true velocity and
acceleration profiles, and the trajectory of the eruption. Numerical modeling for the filament eruption is presented in Paper II
of this series.
Initial results based on SDDI observations can be found in \cite{seki2017, seki2019}, where the increase in the amplitude of the
Doppler velocity is discussed as a possible precursor of filament eruptions. Sun-as-a-star analysis of the SDDI H$\alpha$ spectra
was performed for a number of events by \cite{otsu2022}, and a statistical analysis of flare energies and mass ejections has been
conducted by \cite{kotani2023} recently.

\section{Observations}
\subsection{The Solar Dynamics Doppler Imager (SDDI)}
The Solar Magnetic Activity Research Telescope (SMART) located at the Hida Observatory of Kyoto University consists of four
refractive telescopes \citep[e.g.][]{ueno2004,ishii2013,nagata2014,kawate2016}. SMART/T1 telescope designed to map full-disk solar
images in the H$\alpha$ line had an observing system covering a short set of wavelengths ($\pm 1.2~{\AA}$) around the line center
($6562.8$~{\AA}), which only permitted detections of relatively small Doppler velocity of filament and prominence eruptions. Since
2016, the observing system of SMART/T1 has been upgraded with a new H$\alpha$ imaging instrument named Solar Dynamics Doppler
Imager \citep[SDDI:][]{ichimoto2017}.
This unique instrument consisting of a Lyot type birefringent tunable filter incorporates liquid-crystal variable retarders and is
capable of recording successively high-resolution full-disk solar images at 73~positions (wavelengths) of the H$\alpha$ spectral
profile, spanning $-9$ to $+9$~{\AA} from the line center with a wavelength step of 0.25~{\AA}. This characteristic makes the SDDI
a relevant instrument that allows  to perform not only imaging analysis but also spectroscopy diagnostics of flares and high-speed
mass motions with large Doppler velocities ($\pm400~{\rm~km~s^{-1}}$), and the precise derivation of the velocity vectors in
3D~space. The spatial sampling and time cadence of SDDI is $\sim$$1\farcs23~\rm{pixel}^{-1}$ and $12~\rm{s}$
(for the 73 wavelengths, 15 s before July 2018), respectively, and the field of view covered by the detector of the high-speed CMOS
camera is $2520\times2520$ arcsec${^2}$.

Before proceeding with the analysis of the filament presented in this paper, we performed standard data calibration by applying
dark-current subtraction and flat-field gain corrections. Here it is important to mention that because the flat-pattern is
wavelength-dependent of the tunable filter, we implemented individual flat-field images for each of the SDDI 73 spectral images
following the method of Kuhn-Lin-Loranz \citep{kuhn1991}, see also \cite{ichimoto2017}.

\subsection{Overview of the event}\label{sect_overview}
\begin{figure*}[htbp]  
   \centering                                                                                                               
   \includegraphics[scale=0.9]{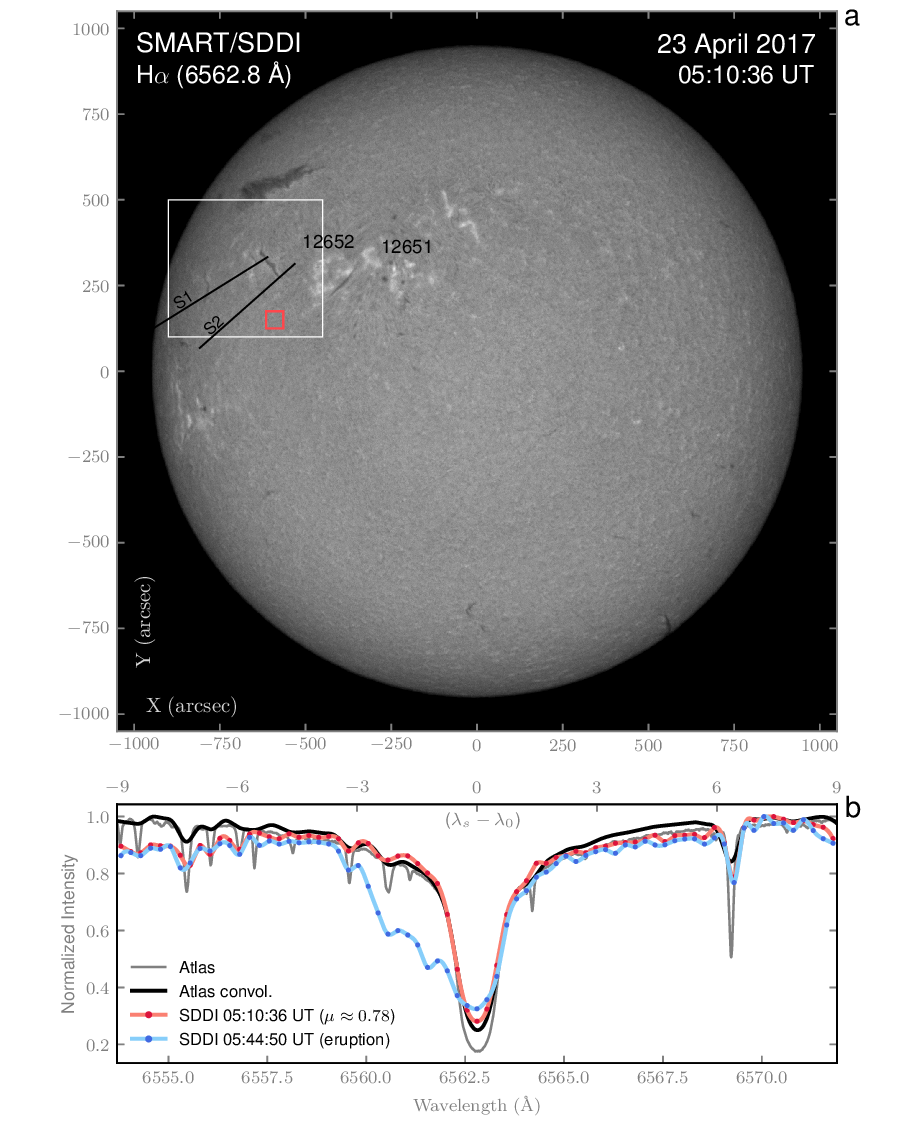}                                                                       
	\caption{Panel (a): solar disk in the H$\alpha$ line (6562.8~\AA) captured by the Solar Dynamics
	Doppler Imager (SDDI) shortly before the filament eruption. The white box encloses the studied                      
        filament and outlines the FOV of the images presented in Fig.~\ref{fig:mosaic}, and the small red box                     
        ($50\arcsec\times50\arcsec$) surrounds a quiet-sun region from which the spectral profile is obtained
	and plotted in panel (b) by the red curve. The slit S1 is positioned in the direction of the leading
	part of the eruption (filament apex), whereas the slit S2 is used to track the lateral part of the                
        eruption.                                                                                                                 
        Panel (b): Atlas solar spectrum of the H$\alpha$ line (gray normalized to the continuum level) and                        
        convolution of the Atlas spectrum with the SDDI filter transmission profile (black). The quiet-sun
	SDDI spectral profile (see above and panel a) at $\mu\approx0.78$ is also shown
	(red dots with spline interpolation), and an example of a strongly blue-shifted profile                             
        (blue dots with spline interpolation) from the lateral part of the filament eruption at 05:44:50~UT                       
        (see Fig.~\ref{fig:mosaic}).                                                                                              
        In the SDDI spectral profiles each dot represents the wavelength offset $\lambda_{s}$ from the line
	center $\lambda_{0}$~($6562.8$~{\AA}) at which SDDI performs regular observations.}                              
     \label{fig:fulldisk}                                                                                                         
\end{figure*}

On 2017 April 23, SDDI detected a filament eruption that occurred in association with a B-class solar flare at the heliographic
coordinates N16~E41 ($X=-605\arcsec$, $Y=335\arcsec$), near NOAA active region 12652. The filament eruption produced a narrow fast
CME traveling with a plane-of-sky velocity of about
$955~{\rm~km~s^{-1}}$.\footnote{https:/cdaw.gsfc.nasa.gov/CME\_list}
The whole process of the filament eruption was captured by SDDI in the full set of wavelengths with a time cadence of 15 seconds.
Figure~\ref{fig:fulldisk} shows a full-disk solar image in the H$\alpha$ line taken at 05:10:36~UT, where the white box encloses
the filament that lies on a quiescent region. In the bottom panel we present the SDDI H$\alpha$ spectral profile of a quiet-sun
region ($X=-580\arcsec$, $Y=148\arcsec$) corresponding to the intensity emerging at $\mu\approx0.78$ (cosine of the heliocentric
angle). The Atlas solar spectrum \citep{kurucz1984} normalized to the continuum level is also shown, as well as the convolution of
the Atlas spectrum with the SDDI filter transmission profile. Also, it is plotted an example of a strongly blue-shifted spectral
profile obtained from the lateral part of the filament eruption at 05:44:50~UT.

\begin{figure*}
   \centering
    \includegraphics[scale=1.1]{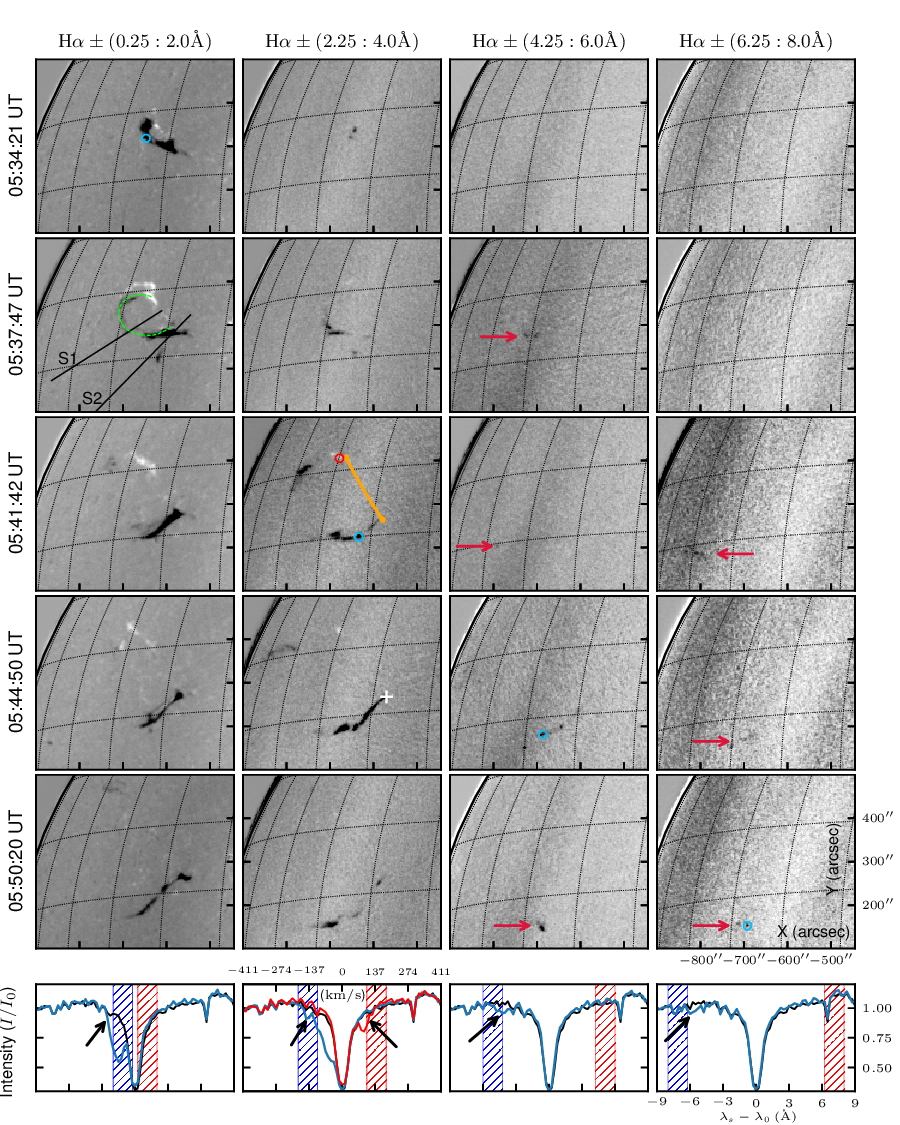}
          \caption{Eruption process of the filament captured in a wide range of the H$\alpha$ spectral line.
	  The maps are Dopplergrams wherein blue- (dark) and red-shifted (bright) features of the erupting
	  material are manifested. The black lines labeled S1 and S2 on the map at 05:37:47~UT represent the
	  artificial slits, and the green curved dashed-line outlines the arc-shaped filament
	  ($\sim$$1.8\times10^{10}~\rm{cm}$). The orange arched line at 05:41:42~UT depicts the filament
	  footpoint separation ($\sim$$1.6\times10^{10}~{\rm cm}$).
	  During the eruption a large amount of the filament material is blue-shifted up to $4.0$~{\AA} from
	  the line center, also blobs-like structures are further blue-shifted over $6.25$~{\AA} (red arrows).
	  The symbol `$+$' at 05:44:50~UT indicates the region from which a strongly blue-shifted spectral
	  profile is obtained and shown in Fig.~\ref{fig:fulldisk}.
          The bottom panels show blue- and red-shifted spectral profiles from the regions outlined with blue
	  and red circle marks (upper panels), the black profiles are from a quiet region. The red-shifted
	  profile (red plot) is caused by downflows close to the filament footpoint (see the text).
	  The background hatched rectangles highlight the wavelength range of the Dopplergrams presented in
	  each column of the upper panels. (An animation of this figure is available)}
      \label{fig:mosaic}
\end{figure*}

Signatures of activation and slow-rising of the filament are observed approximately from 05:10 to 05:25~UT. By 05:30~UT, an
arc-shaped structure already formed and started to grow rapidly becoming these characteristics more noticeable in the blue wing
of H$\alpha$ (e.g. 0.25--2.0~{\AA}). The growing up and rapid expansion of the filament lasted for a few minutes, then at around
05:37~UT the filament apex is fragmented and a large amount of the erupting material is expelled into the outer space
(see Fig.~\ref{fig:mosaic} and the associated animation). What is interesting but not surprising is that the footpoints of the
filament remained rooted to the solar surface and the material contained in the filament legs exhibited signs of untwisting
motions. 
During the late phase of the eruption, that is, from 05:50~UT onward, fragments of the ejected filament fall back onto the solar
surface but a great portion persists ascending, later on, the filament simply fades out and disappears from the H$\alpha$ images
around 06:24~UT. The disappearance of the filament in the H$\alpha$ line suggests that the material possibly reached high
altitudes of the solar atmosphere, also because in the course of the eruption both the increase of the temperature and the
diminution of the density contribute to the ionization \cite[cf.][]{athay1986}, the hot ascending plasma is no longer visible in
cool spectral lines like H$\alpha$.

A time series of the filament eruption captured by SDDI is presented in Fig.~\ref{fig:mosaic}. The images are Dopplergrams
resulting from the difference between mean-intensity maps recorded in the blue and red wings of the H$\alpha$ line, that is,
blue minus red mean-intensity maps in the wavelengths range: H$\alpha\pm(0.25$:$2.0~{\AA})$, H$\alpha\pm(2.25$:$4.0~{\AA})$,
H$\alpha\pm(4.25$:$6.0~{\AA})$, and H$\alpha\pm(6.25$:$8.0~{\AA})$, respectively. The Dopplergrams illustrate the overall
evolution of the filament eruption wherein the material directed towards the Earth is observed as dark structures (blueshift),
while the component moving in the opposite direction is seen as bright structures (redshift).
From 05:41~UT onward, highly blue-shifted material can be identified even at very far-wing wavelengths ($6.25$:$8.0~{\AA}$)
which are indicated by the red arrows (see also the animation). This suggests that the broken parts of the filament are ejected
exceeding a line-of-sight velocity of about $285~{\rm km~s^{-1}}$. A detailed analysis of the highly blue-shifted filament
material caused by the eruption is carried out in Sects. \ref{sect_cloud-model} and \ref{sect_spectral_vel}.
Figure~\ref{fig:mosaic} (panel at 05:37:47~UT) also shows the positions of the artificial slits (S1, S2) used to make the
time-distance diagrams, and the green dashed-line outlines the apparent length of the arc-shaped erupting filament
($\sim$$1.8\times10^{10}~{\rm cm}$). In the second column of Fig.~\ref{fig:mosaic} (panel at 05:41:42~UT) the orange arched line
depicts the filament footpoint separation ($\sim$$1.6\times10^{10}~{\rm cm}$). The footpoint separation and the filament apex
height are important parameters used to characterize the stability and acceleration process of arched magnetic flux ropes
\cite[e.g.][see also Paper II]{chen1989,vrsnak1990,cargill1994,olmedo2010}.
Selected blue- and red-shifted spectral profiles obtained from specific regions of the erupting filament are also shown in the
bottom panels of Fig.~\ref{fig:mosaic}. Here we remark the profile red-shifted to $1.0-2.75~\AA$ (second column) observed at the
filament footpoint. This red-shifted signature could be related to downflows or material precipitating down to the solar surface.
Details of the detected red-shifted features are discussed in Sect.~4.

\section{Analysis and Results}
\subsection{Dynamical evolution and apparent motion}
\begin{figure*}
   \centering
   \includegraphics[scale=0.48]{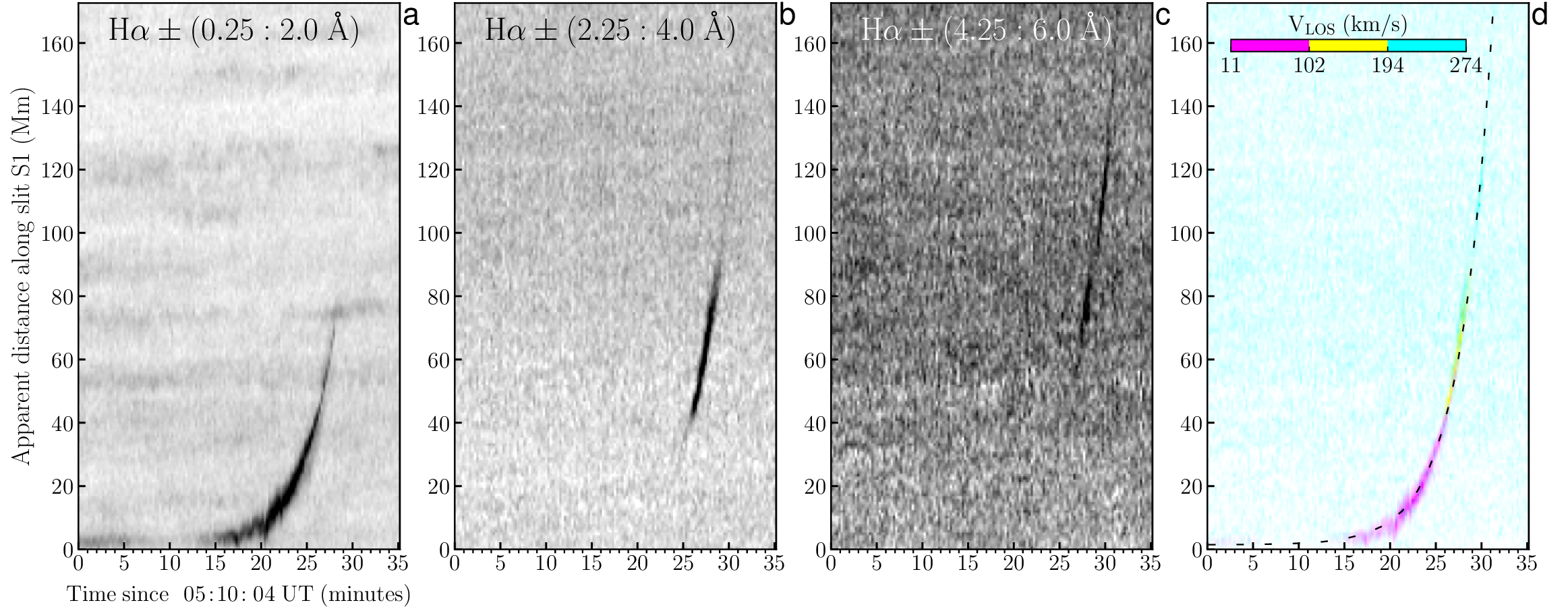}
	  \caption{Multiwavelength time-distance diagrams of the filament eruption obtained along the
	  slit S1 (filament apex). The vertical axes denote the apparent vertical distance in the
	  plane-of-sky measured from the initial position of the filament apex. Panels (a), (b), and (c)
	  show different characteristics of the eruption that become more
	  evident at specific wavelengths. At ${\rm H\alpha}\pm$($0.25$:$2.0$~{\AA}) the slow-rising
	  phase and the initial acceleration are observed, at ${\rm H\alpha}\pm$($2.25$:$4.0$~{\AA})
	  the main phase and fast acceleration are manifested, and at
	  ${\rm H\alpha}\pm$($4.25$:$6.0$~{\AA}) the fastest-ascending phase is visible.
	  Panel (d) is a composite time-distance diagram, which helps to visualize the complete picture
	  of the eruption and demonstrates that the line-of-sight velocity of the erupting filament also
	  increases as a function of distance and time. The inset color-bar depicts the corresponding
	  line-of-sight velocity of the wavelength range indicated in panels (a), (b), and (c). The dashed
	  line is the fitting of an exponential function to the selected points along the erupting feature
	  in the plane-of-sky.
         }
          \label{fig:SDDItimslice}
\end{figure*}

In Sect. \ref{sect_overview} we mentioned that by 05:30~UT the formed arc-shaped filament exhibited a rapid ascending motion
leading to the eruption. The dynamical evolution of the filament can be better visualized as intensity variation in the
multiwavelength time-distance diagrams made from H$\alpha$ imaging spectroscopy observations. Results of the multiwavelength
time-distance diagrams obtained along the filament apex (slit S1), namely, the leading part of the eruption, are presented in
Fig.~\ref{fig:SDDItimslice}. Here it is important to mention that vertical axes of Fig.~\ref{fig:SDDItimslice} denote to the
apparent vertical distance in the plane-of-sky measured along the artificial slit from the initial position of the filament apex.
In principle the diagrams allow to trace the apparent motion of the erupting filament in the plane-of-sky. However, because SDDI
records filtergrams at multiple positions on both sides of the H$\alpha$ line, the obtained diagrams also provide insights of the
line-of-sight component. From the figure it is evident that depending on the wavelength range at which the time-distance plots are
performed, particular characteristics of the erupting material are distinguished. In the diagram H$\alpha\pm(0.25$:$2.0~{\AA})$,
the slow-ascending, the onset, and the main phase of the eruption are observed. At more far-wing wavelengths, that is, in the
diagrams H$\alpha\pm(2.25$:$4.0~{\AA})$ and H$\alpha\pm(4.25$:$6.0~{\AA})$, the fast components of the eruption become visible at
greater distances.
Figure \ref{fig:SDDItimslice} (d) shows a composite time-distance diagram resulting from the combination of the diagrams shown in
panels (a), (b) and (c), that is including observations in the range H$\alpha\pm(0.25$:$6.0~{\AA})$. The composite diagram
illustrates the complete picture of the eruption and demonstrate that as the eruption develops not only the filament apparent
motion (plane-of-sky velocity) increases as a function of distance and time, but also the line-of-sight velocity.
Particularly during the late phase, the erupting filament exhibits a large line-of-sight velocity.
Additionally, an exponential model fitting applied to the measured positions along the erupting feature in the plane-of-sky is
overlaid on the composite diagram. This is to highlight that the evolution of the filament eruption approximates to an exponential
function. The measured positions are also used to estimate the plane-of-sky velocity of the eruption, which in combination with
the line-of-sight velocity (see Sect. \ref{sect_true-profiles}), allows us to derive the true eruption profiles of the filament 
(velocity, height, and acceleration).

\begin{figure*}
   \centering
   \includegraphics[scale=0.68]{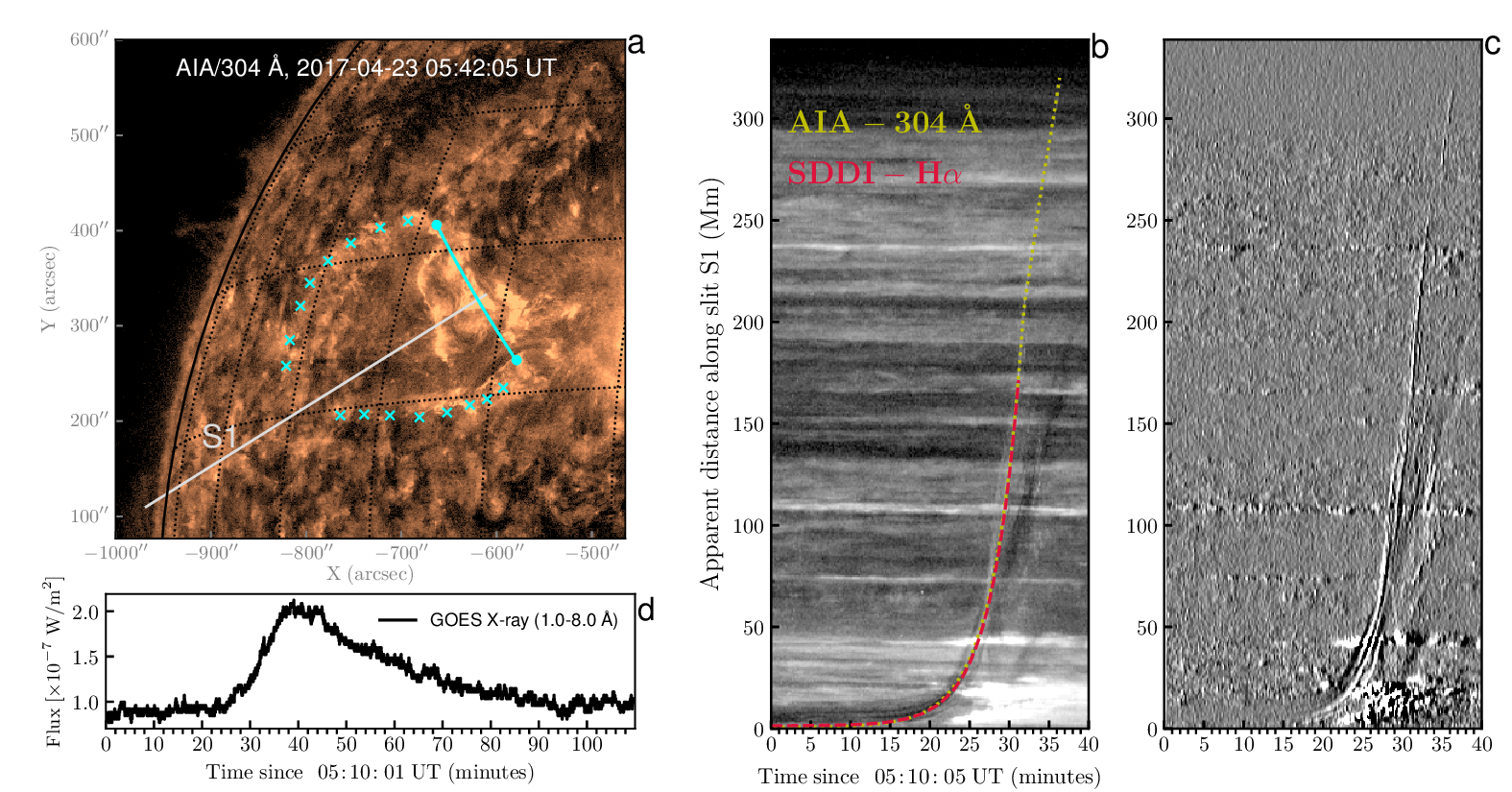}
	  \caption{Panel (a): filament eruption captured by AIA at 304~{\AA} that is dominated
	  by the He~{\sc ii} emission. The straight line labeled S1 outlines the artificial slit
	  positioned along the filament apex (same as for H$\alpha$ in Fig.~\ref{fig:mosaic}).
	  The curved cyan line depicts the filament footpoints separation, while the `$\times$'
	  symbols outline the erupting material.
	  Panel (b): time-distance diagram extracted from the slit S1. The diagram shows a complex
	  eruption, wherein at around 05:38~UT the erupting material splits into two main branches.
	  The yellow dotted profile is the fitting curve applied to the selected positions along
	  the erupting feature in the plane-of-sky, while the red dashed plot is the fitting curve
	  applied to H$\alpha$ data shown in Fig.~\ref{fig:SDDItimslice}.
	  Panel (c): running difference time-distance diagram showing the apparent unwinding motion
	  of the erupting filament (see the text).
	  As in Fig.~\ref{fig:SDDItimslice}, the vertical axis of panels (b) and (c) denote the
	  apparent vertical distance in the plane-of-sky from the initial position of the filament
	  apex.
	  For context, the soft X-ray light curve of the B-class flare is also plotted in panel (d).
	  The temporal evolution of the profiles in panel (b) suggests that the eruption of the
	  filament started before the soft X-ray flux enhancement.}
          \label{fig:AIAtimeslice}
\end{figure*}

For comparison, time-distance diagrams are also obtained from He~{\sc ii} (304~{\AA}) images taken by the Atmospheric Imaging
Assembly \citep[AIA:][]{lemen2012} onboard the Solar Dynamics Observatory \citep{pesnell2012}.
Figure~\ref{fig:AIAtimeslice} shows the filament eruption in the He~{\sc ii} line, where on the map captured at 05:42:05~UT
(panel a) it is depicted the slit S1 (same as for H$\alpha$) from which we extracted the intensity variation. In the He~{\sc ii}
time-distance diagram (panel b) the eruption appears much more complex where two main branches become visible, one evolving ahead
(filament apex) while the other lagging behind. In the diagram the yellow dotted-line represents the fitting of the measured points
along the leading part of the eruption in the plane-of-sky, while the red dashed-line is the fitting result presented in the
H$\alpha$ time-distance diagram (see Fig.~\ref{fig:SDDItimslice}). Here we note that the slow-rise and main phase of the eruption
in the He~{\sc ii} diagram is well fitted with an exponential function, whereas the late phase with a quadratic polynomial form.
By comparing the evolution of the filament eruption in the H$\alpha$ and He~{\sc ii} diagrams, we find that the erupting
features evolve quasi-identically, although the material seen in He~{\sc ii} attains a greater distance ($\sim$$320~{\rm Mm}$) and
is visible for a more extended period. In H$\alpha$ the maximum apparent distance attained by the eruption (filament apex) is
$\sim$$170~{\rm Mm}$, this happens at 05:41~UT instance at which the erupting material practically becomes invisible in the
H$\alpha$ diagram (see Fig.~\ref{fig:SDDItimslice}). On the other hand, the filament eruption also exhibits signs of unwinding
(untwisting) motions. As shown in panel~(c) of Fig.~\ref{fig:AIAtimeslice} such a manifestation is more pronounced during the main
and fast-ascending phases of the eruption.
In the following we will focus on these two phases of the filament eruption, that is, the period in which the eruption is well
observed in H$\alpha$. This is also because we aim to calculate the true eruption profiles using imaging spectroscopy observations
(see Sect. \ref{sect_true-profiles}).


\subsection{Doppler velocity based on the ``cloud model"}\label{sect_cloud-model}
The classical ``cloud model'' introduced by \citet{beckers1964} is an effective tool to derive the physical properties of
filaments and prominences, as well as of fine structures in the solar chromosphere. In the model such structures are considered
as plasma ``clouds" located above the solar surface, which absorb and scatter the incident radiation coming from the solar surface
behind. The process in which the incident radiation passes through and emerges from the cloud structure is well described by the
radiative transfer equation. The simplified form of the solution of the radiative transfer equation is:

\begin{equation}
        I(\Delta\lambda) = I_{0}(\Delta\lambda)e^{-\tau(\Delta\lambda)}+ S(1- e^{-\tau(\Delta\lambda)}),
\label{Eq:cmodel}
\end{equation}
where $I(\Delta\lambda)$ represents the observed intensity, $I_{0}(\Delta\lambda)$ the background incident radiation,
$\tau(\Delta\lambda)$ the optical thickness, and $S$ the source function. Equation~(\ref{Eq:cmodel}) arises from the assumption
that the cloud structure is fully isolated from the underlying surface, the source function, Doppler width, and the absorption
coefficient are constant along the line-of-sight \cite[see][for comprehensive reviews]{tziotziou2007,heinzel2015}.
From Eq.~(\ref{Eq:cmodel}) the contrast profile $C(\Delta\lambda)$ is defined as:

\begin{equation}
        C(\Delta\lambda) = \frac{I(\Delta\lambda) - I_{0}(\Delta\lambda)}{I_{0}(\Delta\lambda)}
        = \left(\frac{S}{I_{0}(\Delta\lambda)}-1\right) (1-e^{-\tau(\Delta\lambda)}).
\label{Eq:contrast}
\end{equation}
Note that the variables on the left side of Eq.~(\ref{Eq:contrast}) can independently be determined from the observations.
Furthermore, for practical reasons it is assumed that the optical thickness of the cloud follows a Gaussian profile:

\begin{equation}
        \tau(\Delta\lambda) = \tau_{0}e^{-\left(\frac{\Delta\lambda-\Delta\lambda_{los}}{\Delta\lambda_{D}} \right)^{2}},
\label{Eq:tau1}
\end{equation}
here $\tau_{0}$ is the optical thickness at the line center, $\Delta\lambda=\lambda-\lambda_0$ is the wavelength displacement from
the H$\alpha$ line core $\lambda_{0}$, $\Delta\lambda_{D}$ the Doppler width, and
$\Delta\lambda_{los}=\lambda_{0} \upsilon_{los}/c$ is the Doppler shift of the line due to the line-of-sight motion of the cloud
$\upsilon_{los}$ being $c$ the speed of light.

\begin{figure*}[!h]
   \centering
   \includegraphics[scale=1.0]{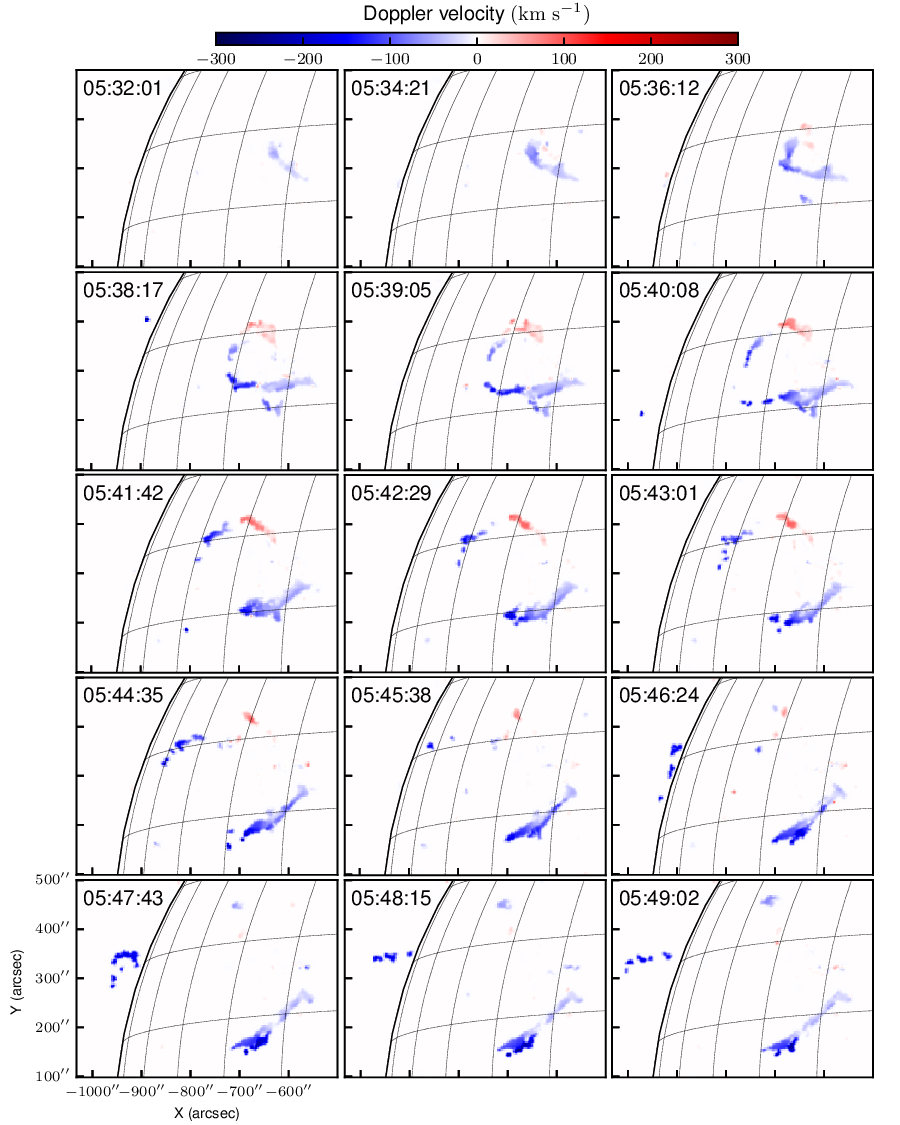}
        \caption{Velocity maps based on the ``cloud model'' illustrating the Doppler characteristics of the
	filament eruption (blue colors toward the Earth). Shortly after the filament disruption, which happens
	at around 05:38~UT, a large amount of the filament material is launched with velocities
	of about $-100~{\rm km~s^{-1}}$ (blue-shifted).
	Red-shifted structures corresponding to the material moving in opposite direction are also observed
	close to the footpoints of the erupting filament. This suggests that downflows are driven during the
	eruption and could be related to the mass loss (see the text). In the latter phase of the eruption,
	blob-like structures traveling with velocities exceeding $-250~{\rm km~s^{-1}}$ also can be identified.
	(An animation of this figure is available)}
       \label{fig:VLOS_maps}
 \end{figure*}

Applying the above definitions to the SDDI imaging spectroscopy data, we obtained a series of two-dimensional physical properties
of the filament eruption including the Doppler shift relative to the line center. From the Doppler shift we directly derived
Doppler velocity maps that are presented in Fig.~\ref{fig:VLOS_maps}.
Following the convention, negative Doppler velocity corresponds to blueshift (towards the observer), while positive Doppler
velocity to redshift (away from the observer).
The velocity maps illustrate in great detail the Doppler characteristics of the filament eruption, where the color-code highlights
the velocity distribution of the plasma contained in the ejecta, that is, red- and blue-shifted patterns in relation to the
observer. During the initial phase of the eruption, from 05:20 to 05:34~UT (see the associated animation), it is noted a moderated
increase of the Doppler velocity (blueshift) up to several tens of kilometers~per~second. Subsequently, there is a drastic
increment in the velocity due to the strong blue-shifted plasma.
The most remarkable change in the filament configuration and its velocity is visualized from 05:38:17~UT, time when the filament
apex is fragmented and the material that is being erupted starts to exhibit red- and particularly highly blue-shifted
characteristics. During this period, which corresponds to the main phase of the eruption, a great amount of the filament material
is launched with a line-of-sight velocity larger than $-100~{\rm km~s^{-1}}$. Red-shifted structures also appear predominantly in
the filament upper leg, which suggests that downflows are manifested as the filament expands upward.
Shortly after, from 05:42 to 05:49~UT blob-like structures are ejected even with larger velocities, most of the blobs approaching
$-300~{\rm km~s^{-1}}$ (see the animation of Fig.~\ref{fig:VLOS_maps}). The characteristics of large Doppler velocity detected
during the filament eruption are further investigated in the next section.

\subsection{Spectral diagnostic and multiple velocity components}\label{sect_spectral_vel}
Here we investigate the spectral characteristics of the filament eruption focusing on the highly Doppler-shifted material detected
at the apex and the lateral part of the eruption. To this end, we take advantage of the SDDI high-resolution full-disk spectral
images that allow us to obtain two-dimensional spectra and profiles from any given position and extent on the solar disk.

In order to get a more physically meaningful quantity from our observation, we rewrite Eq.~(\ref{Eq:contrast}) as: 

\begin{equation}
	\tau(\Delta\lambda) = -\ln\left(1+ \frac{I(\Delta\lambda)-I_{0}(\Delta\lambda)}{I_{0}(\Delta\lambda)-S}\right).
\label{Eq:tau2}
\end{equation}
Thus, using  Eq.~(\ref{Eq:tau2}) the optical thickness can be calculated. Except for the source function $S$ all quantities in
Eq.~(\ref{Eq:tau2}) can be inferred from the observation. We use two-dimensional spectra as inputs for $I_{0}(\Delta\lambda)$ and
$I(\Delta\lambda)$ obtained along the slits S1 (filament apex) and S2 (lateral part of the eruption), respectively. Post-eruption
spectra at 06:10:03~UT along the corresponding slits are considered as background radiation $I_{0}(\Delta\lambda)$, and we set
$S=0.1$ expressed in units of the disk-center continuum intensity around the H$\alpha$ line 
(${\rm 4.077\times10^{-5}~erg~s^{-1}~cm^{-2}~sr^{-2}~Hz^{-1}}$).
It should be mentioned that increasing or decreasing $S$ in Eq.~(\ref{Eq:tau2}) produces minimal effect in the final result, for
instance, only a small variation in the amplitude if one calculates optical~thickness profiles (see discussion
Sect.~\ref{Doppler_components}).
In this way, we computed two-dimensional optical thickness spectra in which the erupting material can be traced more effectively
in space (slit), time $(t)$, and wavelength $(\Delta\lambda)$.
Furthermore, we applied deconvolution to the optical thickness spectra with the theoretical transmission profile of the Lyot filter
to recover the true optical thickness profiles.
Thereafter, we reconstructed a series of optical thickness spectra (distance along the slit and $\Delta\lambda$) by assembling the
deconvolved profiles. 

\begin{figure*}
   \centering
   \includegraphics[scale=0.70]{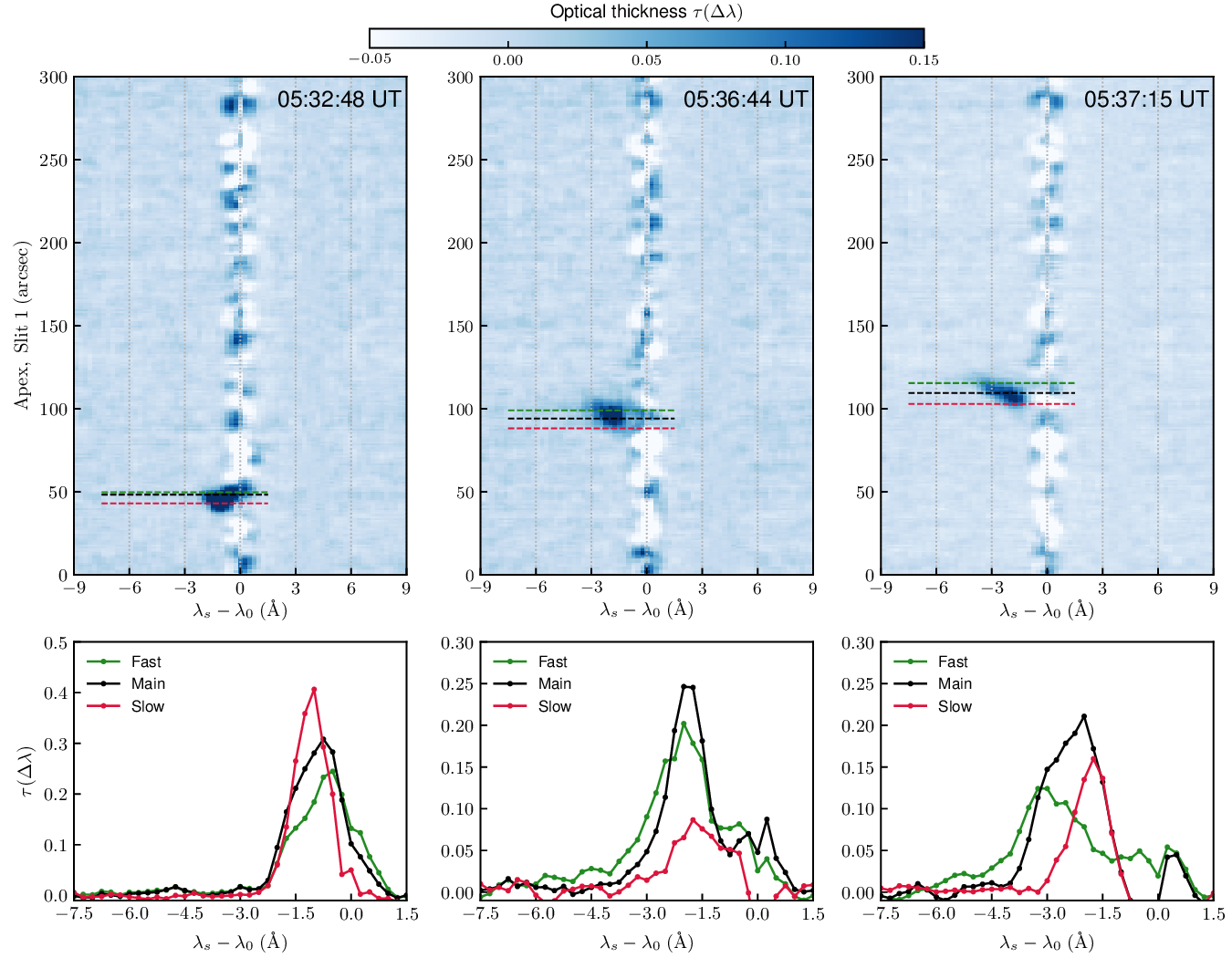}
	\caption{Upper panels: optical thickness spectra of the filament apex (slit S1) extracted from the SDDI
	H$\alpha$ filtergrams (see Fig.~\ref{fig:mosaic}). We applied deconvolution to recover the instrumental
	broadening in our observations (see the text). In the two-dimensional spectra, the erupting filament is
	visible as dark elongated structures extending upward from the central part of the spectra. The horizontal
	green, black, and red lines are used to track the fast (leading), main (core), and slow (following)
	components of the eruption, respectively.
	We note that the negative values in the optical thickness spectra are related to some errors (bias) in
	the estimates arising from the noise in the data.
        Lower panels: optical thickness profiles obtained from the horizontal lines shown in the upper panels.
	The profiles show different characteristics of the erupting material, for example, multiple peaks and large
	Doppler shifts. (An animation of this figure is available)}
        \label{fig:tau2d_apex}
\end{figure*}

\begin{figure*}
   \centering
   \includegraphics[scale=0.70]{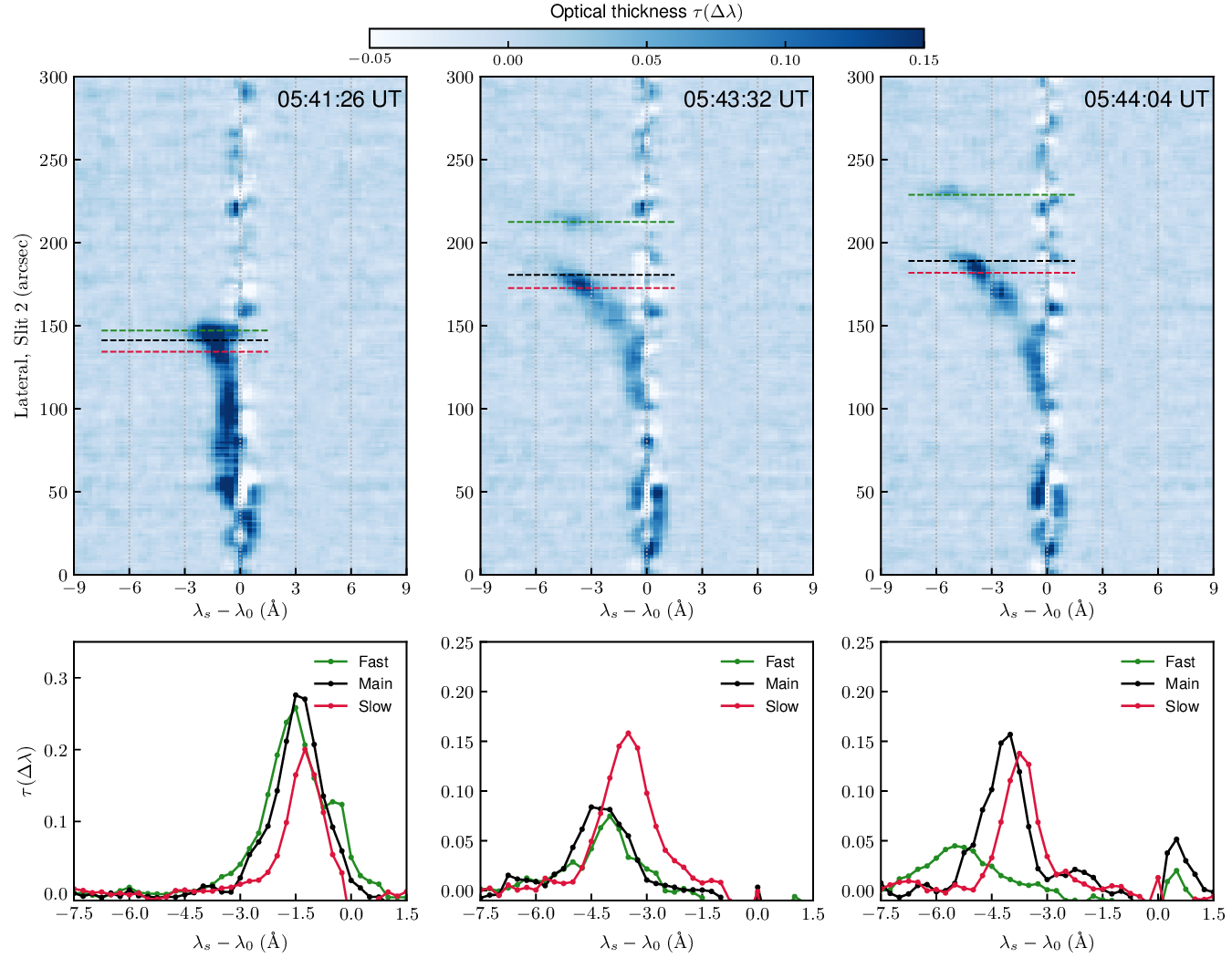}
        \caption{Same as in Fig.~\ref{fig:tau2d_apex} but for the lateral part of the filament eruption that
	crosses the slit~S2 (see Fig.\ref{fig:mosaic}). What is remarkable here is the detection of strongly
	blue-shifted plasma contained in the filament. Again, the optical thickness profiles help to visualize
	in more detail the characteristics of the fast, main, and slow components of the erupting material.
	(An animation of this figure is available)}
        \label{fig:tau2d_lat}
\end{figure*}

Examples of two-dimensional optical~thickness resulting from deconvolution are presented in Figs.~\ref{fig:tau2d_apex} and
\ref{fig:tau2d_lat} (upper panels) for the slits S1 and S2, respectively. In the optical thickness spectra the erupting material
is observed as dark elongated structures departing from the line center (see the associated animations). These characteristics
indicate that the eruption is made up predominantly of blue-shifted plasma. In the course of time, the dark structures become
faint which suggests that the erupting material reached high altitudes of the solar atmosphere, it started to get ionized due to
the temperature enhancement and/or density reduction.
Because in the two-dimensional spectral image the highly blue-shifted plasma can be tracked more effectively, optical thickness
profiles are obtained at three positions across the erupting material, namely, fast (leading), main (core), and slow (following)
components (Figs.~\ref{fig:tau2d_apex}, \ref{fig:tau2d_lat} lower panels). The profiles help to visualize more characteristics of
the erupting material, for example, the profiles have no symmetric shapes nor single peaks. In addition to the main blue-shifted
peak, there exist multiple peaks or sub-components that are manifested exhibiting different behaviors during the filament eruption. 

\begin{figure*}[!h]
  \centering
  \includegraphics[scale=0.8]{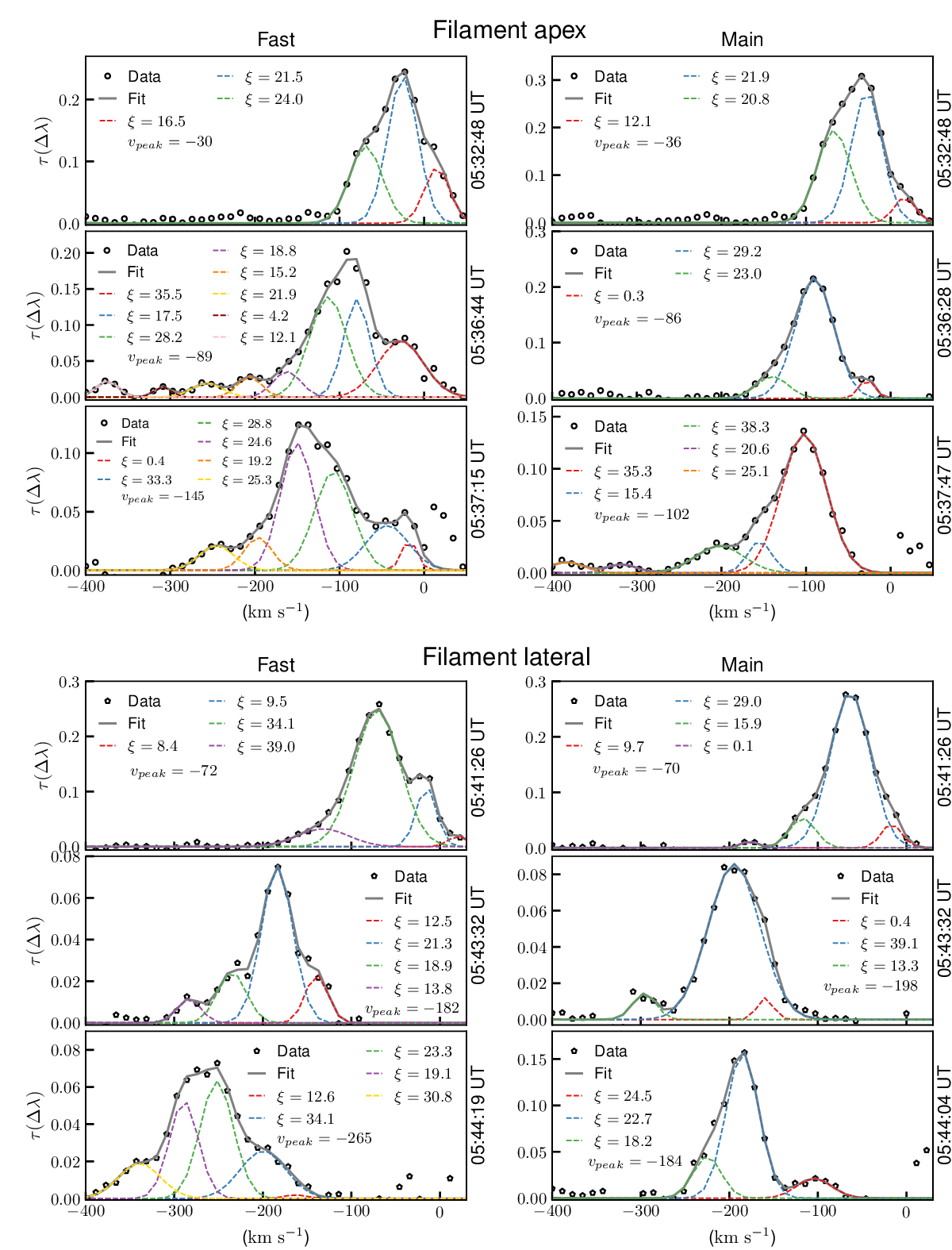}
        \caption{Selected optical thickness profiles of the fast and main components at the apex (top panels)
	and lateral part (bottom panels) of the filament eruption, respectively. The time of the profiles are
	nearly the same as the two-dimensional spectra shown in Figs.~\ref{fig:tau2d_apex} and
	\ref{fig:tau2d_lat}. The colored profiles highlight the sub-components resulting from the Gaussian
	fitting model, while the solid gray profiles are composite fitting results by combining the multiple
	sub-components. It is also shown the peak velocity $\upsilon_{peak}~(\rm km~s^{-1})$ of the composite
	profiles and the estimated micro-turbulent velocity $\xi~(\rm km~s^{-1})$ of each sub-component
	(see the text).}
        \label{fig:ApexLatFastMain}
\end{figure*}

We further characterize the optical thickness profiles with multiple Gaussian fits by employing Eq.~(\ref{Eq:tau1}) as a model
fitting. The Doppler width in Eq.~(\ref{Eq:tau1}) is related to the mass of the hydrogen atom
$m_H\simeq1.67\times10^{-24}~{\rm g}$ and the unknown kinetic temperature $T$ and micro-turbulent velocity $\xi$ due to the
non-thermal motions. Given that the Doppler velocity $\upsilon_{los}$ is a measurable quantity from our observation and assuming
that the temperature of the erupting filament (cloud) is about $T\approx10^4~{\rm K}$, we created synthetic optical thickness
profiles by applying the model fitting. The fitting results provide us with better estimates of the model coefficients, such as
the peak velocity $\upsilon_{peak}$, the optical thickness at the line center $\tau_0$, and $\xi$ for each synthetic profile. The
combination of the synthetic profiles yields the composite fitting curves for the observed optical thickness profiles.
In Fig.~\ref{fig:ApexLatFastMain} we present selected optical thickness profiles of the filament eruption (apex and the lateral
part) along with the fitting results of the multiple Gaussian model. It is also shown the individual synthetic profiles to
highlight the multi-component characteristics of the filament eruption. The profiles clearly show that the erupting filament has
a main peak velocity and multiple sub-components, some profiles exhibit strong blueshift denoting the presence of large Doppler
velocity. This means that during the eruption there exists a main bulk moving with an average velocity determined by the peak of
the distribution and the other sub-components or unresolved structures moving ahead and behind the main bulk with distinct
velocities. The multiple peaks that appear in the profiles could be related to the presence of internal motions inside the
erupting filament (see discussion in Sect. \ref{sect_discussion}).

\subsection{Velocity distribution and true eruption profiles}\label{sect_true-profiles}
\begin{figure*}
   \centering
   \includegraphics[scale=0.75]{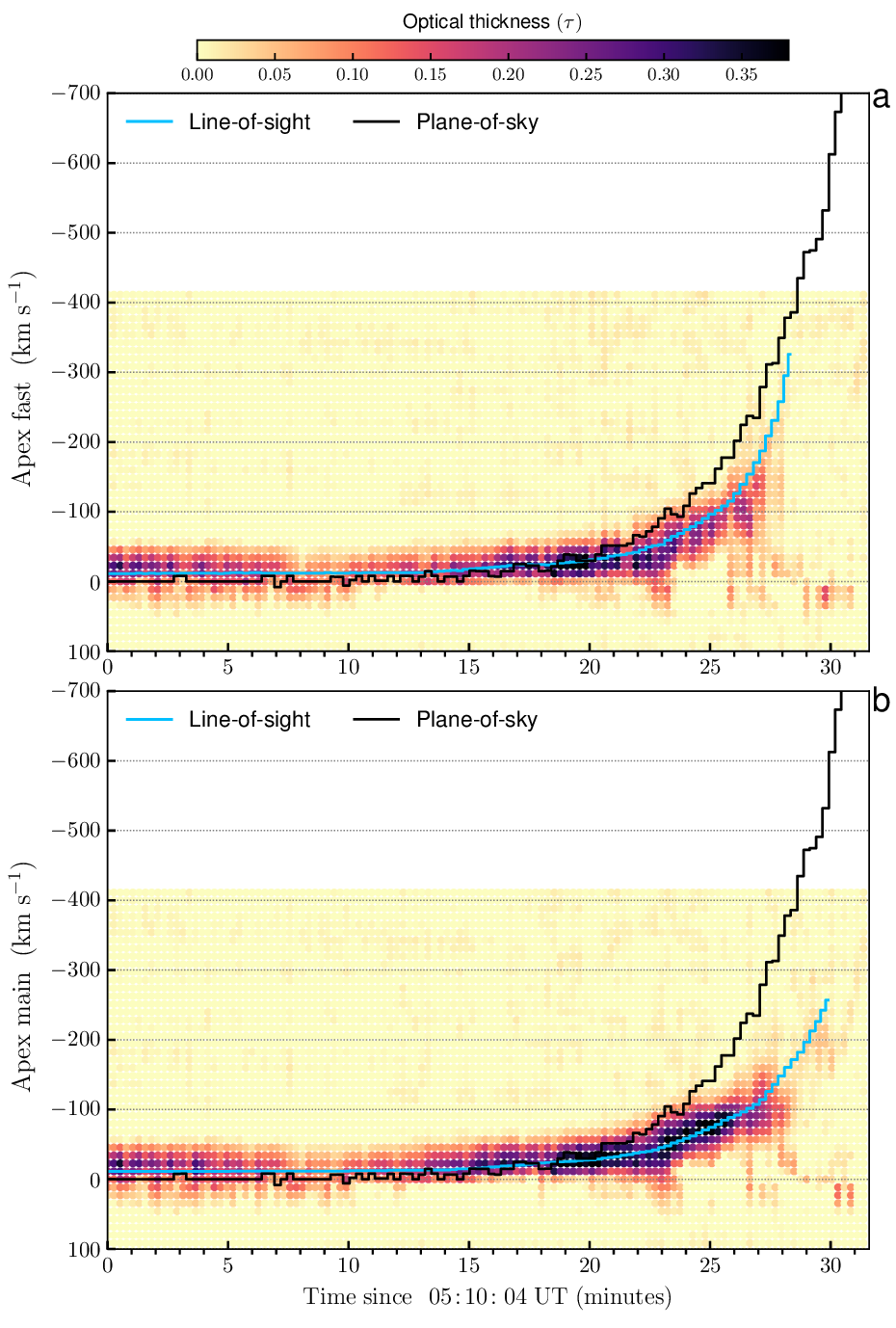}
        \caption{Distribution of the optical thickness profiles obtained from the filament apex. Panel (a)
	shows the distribution of the fast component, while panel (b) the distribution of the main component.
	The fast and main components refer to the measurements obtained from the leading and the central part
	of the erupting feature, respectively, detected in the optical thickness spectra (see the text). The
	enhanced regions correspond to the multiple peaks of the profiles which are Doppler-shifted to large
	velocities during the filament eruption.
	The blue profiles are the line-of-sight velocity $\upsilon_{los}$ of the fast and main components
	calculated by tracking the enhanced regions. For comparison, it is also plotted the plane-of-sky
	velocity profile $\upsilon_{pos}$ (black) calculated from the time-distance diagrams presented in
	Fig.~\ref{fig:SDDItimslice}.
         }
        \label{fig:Vlos_apex}
\end{figure*}

The optical thickness profiles described in the previous section are also used to infer the distribution of the line-of-sight
velocity. Because the optical thickness is a function of wavelength $\tau(\Delta\lambda)$, it can conveniently be expressed in
terms of velocity. Figure \ref{fig:Vlos_apex} presents the distribution and time-variation of the optical thickness profiles
corresponding to the fast and main components at the apex of the erupting filament. By fast and main components we mean the
profiles obtained from the leading and the central part of the erupting feature identified in a series of the deconvolved
two-dimensional optical thickness spectra (e.g. Fig.~\ref{fig:tau2d_apex}).
In Fig.~\ref{fig:Vlos_apex} the concentrated points seen as enhanced regions correspond to the highest peaks of the optical
thickness profiles. The distribution plots illustrate that during the initial period highest values of the optical thickness are
congregated, on average, below $-10~{\rm km~s^{-1}}$, indicating a small Doppler shift. From 05:25~UT the peak values of the
optical thickness are blue-shifted to relatively large velocities, and after 05:35~UT the peaks of the optical thickness start to
get shifted even to larger velocities and some peaks approaching $-$$300~{\rm km~s^{-1}}$ in the latter phase. Starting from
05:37~UT onward, it can be seen large velocity dispersion, suggesting the existence of multi-components that are predominantly
manifested during the final episode of the filament eruption.
Figure~\ref{fig:Vlos_Lat} presents the optical thickness distribution of the lateral part of the filament eruption also obtained
from a sequence of two-dimensional spectra as shown in Fig.~\ref{fig:tau2d_lat}.

\begin{figure}
   \centering                                                                                                                    
   \includegraphics[scale=0.56]{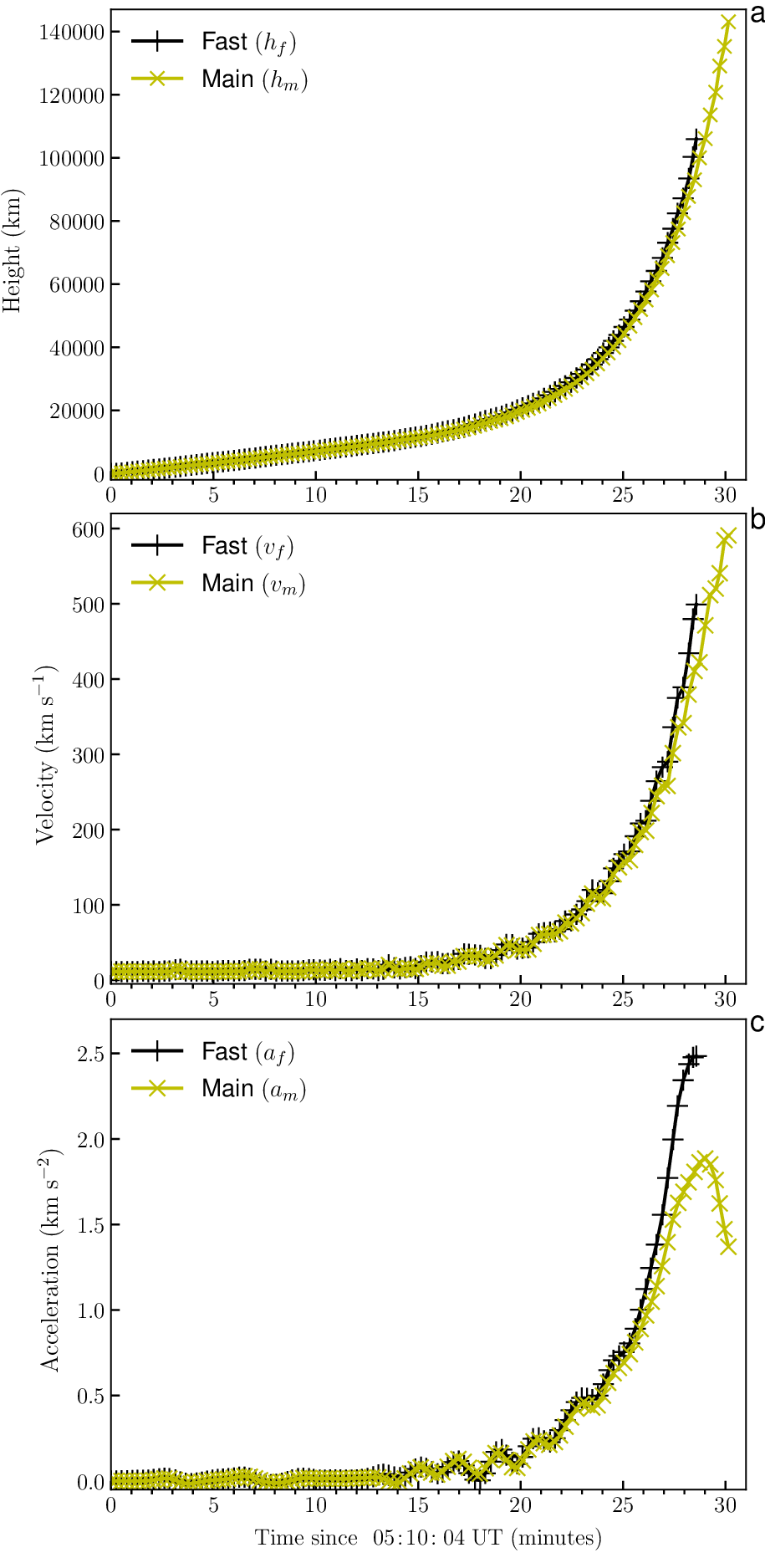}        
        \caption{True eruption profiles (height, velocity, and acceleration) of the filament apex fast
	and main components. We note that the profiles in panel (a) show the actual propagation distance
	of the filament eruption measured from the initial position of the filament apex (see the text).
	The eruption profiles reveal that the filament attains a large propagation distance (panel a),
	high propagation speed (panel b), and fast acceleration (panel c). Of note regarding the
	velocity and acceleration profiles is that the fast and main components evolve identically
	until 05:35~UT, thereafter the profiles deviate from each other, that is, one accelerates
	impulsively while the other undergoes a rapid diminution. 
        }                                                                                                                         
        \label{fig:true_hva}                                                                                                      
\end{figure}

Given the optical thickness distribution, we calculate the line-of-sight velocity profiles of the fast and main components by
approximately tracking at each time~step the central part of the congregated optical thickness peaks. We note that in the optical
thickness distribution (Fig.~\ref{fig:Vlos_apex}) evidence of larger Doppler shifts does exist, identified as weak enhancements
at far distances from the central part. In our analysis we consider the derived velocity profiles as representative line-of-sight
components $\upsilon_{los}$, which in combination with the plane-of-sky component $\upsilon_{pos}$ obtained nearly at the same
location along the slit S1, the true velocity at the apex of the erupting filament is calculated as
$\upsilon_{true} = \sqrt{\upsilon_{los}^2 + \upsilon_{pos}^2}$.
By integrating and differentiating the velocity profiles we calculate the true height and acceleration profiles, respectively.
It is noteworthy here that the true height refers to the actual propagation distance of the filament eruption measured along the
slit S1 from the initial position of the filament apex. Figure~\ref{fig:true_hva} presents the true eruption profiles of the fast
($h_f$, $\upsilon_f$, $a_f$) and main ($h_m$, $\upsilon_m$, $a_m$) components. The eruption profiles show interesting
characteristics of the filament dynamics, for example, a large propagation distance, high propagation speed, and fast acceleration.
These aspects make the filament eruption under study a relevant phenomenon that is not often observed. Comparing the velocity and
acceleration profiles of the main and fast components, we observe that they evolve almost identically during the initial and
ascending phase of the eruption until about 05:35~UT. Subsequently, the profiles deviate from each other, that is, the fast
component accelerates impulsively while the main component experiences a gradual increase and then it steeply drops. These findings
are further investigated by performing numerical simulations in Paper II of this series.

\subsection{Propagation direction of the filament}
\begin{figure*}[ht] 
   \centering                                                                                                                    
\includegraphics[width=14cm]{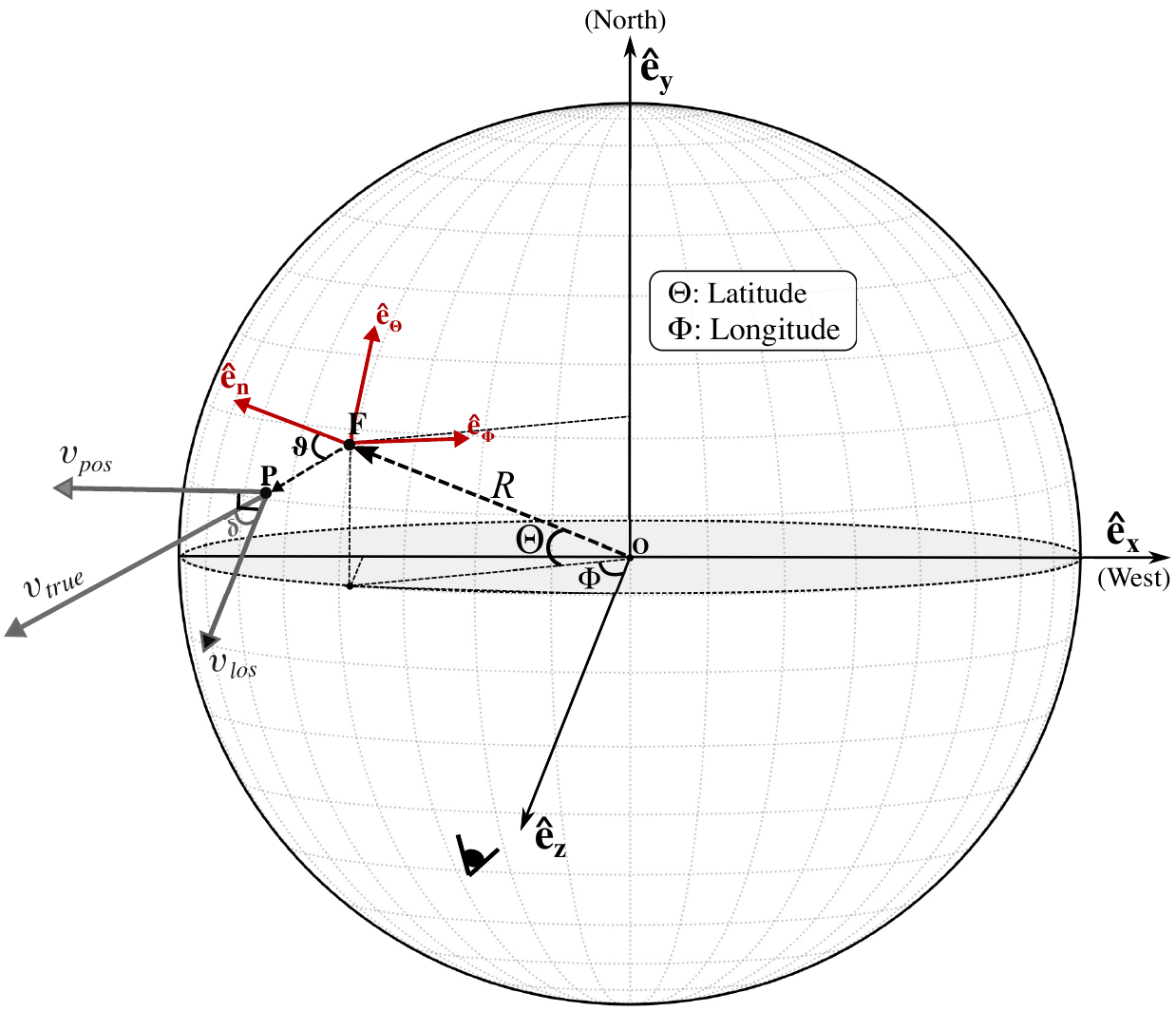}  
        \caption{Geometry of the Sun and representation of the velocity vectors related to the
	filament eruption. The axes ($\mathbf{\hat{e}_{x}},\mathbf{\hat{e}_{y}},\mathbf{\hat{e}_{z}}$)
	define the Cartesian coordinate system (observer's reference frame), while
	($\mathbf{\hat{e}_{\Phi},\hat{e}_{\Theta},\hat{e}_{n}}$) determine the local coordinate system
	(local reference frame) fixed at point $\mathbf{F}$ (see the text). $\mathbf{P}$ indicates an
	instantaneous position of the filament used to describe the trajectory of the eruption in 3D
	space, which is  presented in Fig.~\ref{fig:inc_3D}.
	The arrows labeled $\upsilon_{pos}$, $\upsilon_{los}$, and $\upsilon_{true}$ denote the
	plane-of-sky, line-of-sight, and true velocities, respectively. The angle of the eruption
	subtended to the Sun-Earth line is given by $\delta$, whereas the inclination angle of the
	eruption with respect to the local solar normal $\mathbf{\hat{e}_{n}}$ is denoted by
	$\vartheta$.
        }                                                                                                                         
\label{fig:inc_coords}                                                                                                            
\end{figure*}

The propagation direction of the filament eruption in 3D space is examined here by considering the velocity vectors derived from
our observation.
Figure~\ref{fig:inc_coords} shows a schematic representation of the Sun, where ($\mathbf{\hat{e}_x}$, $\mathbf{\hat{e}_y}$,
$\mathbf{\hat{e}_z}$) defines the right-handed Cartesian coordinate system (observer's reference frame) with the axes centered at
the Sun's center and orthogonal to each other. The $\mathbf{\hat{e}_z}$ axis is aligned with the line-of-sight
direction, while $\mathbf{\hat{e}_x}$ and $\mathbf{\hat{e}_y}$ span the plane-of-sky with $\mathbf{\hat{e}_x}$ pointing towards
the solar West and $\mathbf{\hat{e}_y}$ towards the solar North. The vector $\mathbf{OF}$ denotes the position of the filament
before its eruption, $\Theta=16\deg$ and $\Phi=-41\deg$ are the heliographic latitude and longitude of $\mathbf{F}$, respectively.
On the other hand, an instantaneous position of the erupting filament is denoted by $\mathbf{P}$ and its velocity components in
the plane-of-sky $\upsilon_{pos}$ and in the line-of-sight $\upsilon_{los}$, and the true velocity $\upsilon_{true}$ are
represented by the gray arrows.
The propagation direction of the eruption relative to the Sun-Earth line is given by the angle $\delta$, while $\vartheta$ is the
angle that the eruption makes with the local solar normal $\mathbf{\hat{e}_n}$ (see below). Based on the geometric relations these
angles are given by: $\tan\delta = \frac{\upsilon_{pos}}{\upsilon_{los}}$ and
$\cos\vartheta = \mathbf{\hat{e}_n}\cdot\mathbf{\hat{e}_{true}}$, where $\mathbf{\hat{e}_{true}}$ refers to the unit vector
parallel and in the direction of the true velocity.

To calculate the three-dimensional trajectory of the filament eruption with respect to the solar surface, we define a local
coordinate system (local reference frame) determined by the unit vectors ($\mathbf{\hat{e}_{\Phi}}$, $\mathbf{\hat{e}_{\Theta}}$,
$\mathbf{\hat{e}_n}$) fixed at point $\mathbf {F}$ (see Fig.~\ref{fig:inc_coords}). Here $\mathbf{\hat{e}_n}$ is a normal vector
perpendicular to the plane tangential to the solar surface and, $\mathbf{\hat{e}_{\Phi}}$ and $\mathbf{\hat{e}_{\Theta}}$ are
tangent vectors directed toward the solar West and North, respectively.
The unit vectors ($\mathbf{\hat{e}_{\Phi}}, \mathbf{\hat{e}_{\Theta}}, \mathbf{\hat{e}_n}$) are calculated as follows. From the
geometry the position vector $\mathbf{OF}$ is written as:

\begin{equation}
        \mathbf{OF} = R\cos\Theta\sin\Phi\mathbf{\hat{e}_x} + R\sin\Theta\mathbf{\hat{e}_y} +
                R\cos\Theta\cos\Phi\mathbf{\hat{e}_z},
\label{Eq:OF}
\end{equation}
where $R$ represents the solar radius. Since the position vector $\mathbf{OF}$ is parallel to the unit vector $\mathbf{\hat{e}_n}$,
meaning that both are perpendicular to the tangential plane at point $\mathbf {F}$, we find that
$\mathbf{\hat{e}_{n}} = \frac{\mathbf{OF}}{|\mathbf{OF}|}$. Moreover, from the geometrical relations and calculating the
cross-product, $\mathbf{\hat{e}_{\Phi}}$ and $\mathbf{\hat{e}_{\Theta}}$ can be easily derived. Therefore, the unit vectors
belonging to the local reference frame are given by:
\begin{equation} 
	\mathbf{\hat{e}_{\Phi}} = \cos\Phi\mathbf{\hat{e}_x} - \sin\Phi\mathbf{\hat{e}_z},
\label{Eq:e_phi}                
\end{equation}
\begin{equation}
	\mathbf{\hat{e}_{\Theta}} = -\sin\Theta\sin\Phi\mathbf{\hat{e}_x} +
        \cos\Theta\mathbf{\hat{e}_y} - \sin\Theta\cos\Phi\mathbf{\hat{e}_z},
\label{Eq:e_theta}
\end{equation}
\begin{equation}                                                                                                                  
        \mathbf{\hat{e}_n} = \cos\Theta\sin\Phi\mathbf{\hat{e}_x} + \sin\Theta\mathbf{\hat{e}_y}                                  
        + \cos\Theta\cos\Phi\mathbf{\hat{e}_z}.
\label{Eq:en}                                                      
\end{equation} 
Next, we define the position vector of the erupting filament with respect to point $\mathbf{F}$ to describe the trajectory of the
eruption. This vector can be written in terms of the observer's and the local reference frames, respectively, as:

\begin{equation}
	\mathbf {FP} = X(t)\mathbf{\hat{e}_x} + Y(t)\mathbf{\hat{e}_y} + Z(t)\mathbf{\hat{e}_z},
\label{Eq:FP1}
\end{equation}

\begin{equation}                                                                                                                  
	\mathbf {FP} = \alpha(t)\mathbf{\hat{e}_{\Phi}} + \beta(t)\mathbf{\hat{e}_{\Theta}} + \gamma(t)\mathbf{\hat{e}_n},
\label{Eq:FP2}                                                                                                                          
\end{equation} 
where ($X(t),Y(t),Z(t)$) refer to the coordinates in the observer's reference frame, while ($\alpha(t),\beta(t),\gamma(t)$) the
coordinates in the local reference frame.
We can relate Eqs. (\ref{Eq:FP1}) and (\ref{Eq:FP2}) through the following transformation matrix equation:
\begin{equation}
	\begin{pmatrix}
	\alpha \\
	\beta \\
	\gamma
	\end{pmatrix}_{t}
	=
	\begin{pmatrix}
	\cos\Phi            &0          &-\sin\Phi \\
	-\sin\Theta\sin\Phi  &\cos\Theta &-\sin\Theta\cos\Phi \\
	\cos\Theta\sin\Phi &\sin\Theta &\cos\Theta\cos\Phi
	\end{pmatrix}
	\begin{pmatrix}
	X(t) \\
	Y(t) \\
	Z(t)
	\end{pmatrix}.
\label{Eq:matrix}
\end{equation}
Equation~(\ref{Eq:matrix}) allows the conversion of the observer's coordinate into the local coordinate system. We note that
$X(t)$ and $Y(t)$ are the positions of the filament apex in the plane-of-sky, while it is very close to the slit S1 as a result.
$Z(t)$ is obtained from the time integral of $\upsilon_{los}(t)$ profile of the apex main component.

\begin{figure*}
   \centering
   \includegraphics[scale=0.67]{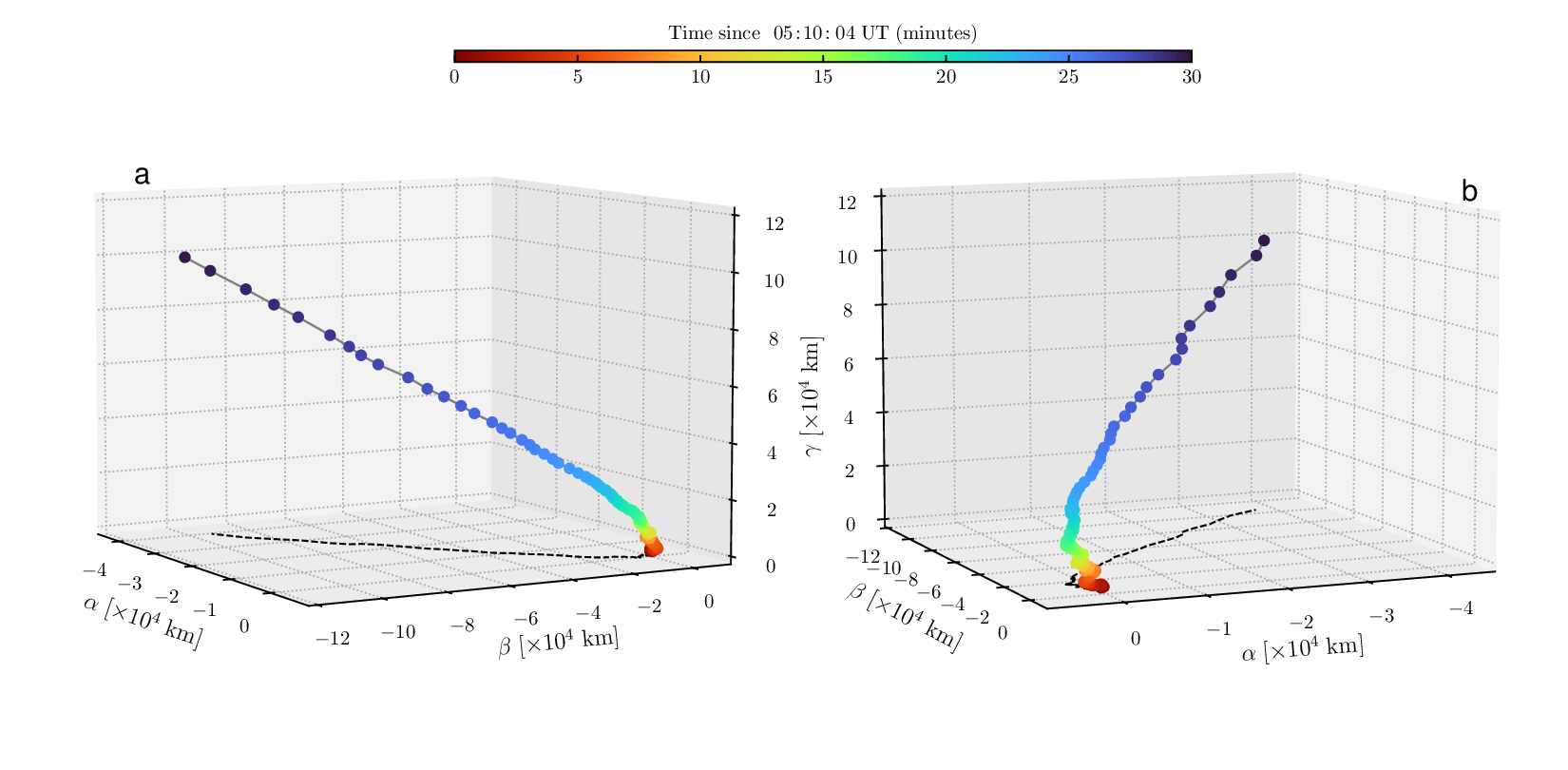}
	\caption{
	Trajectory of the filament eruption projected in 3D space calculated with Eq.~(\ref{Eq:matrix}).
	The axes ($\alpha,\beta,\gamma$) refer to the local coordinates system determined by the unit
	vectors ($\mathbf{\hat{e}_{\Phi},\hat{e}_{\Theta},\hat{e}_{n}}$) (see Fig.~\ref{fig:inc_coords}).
	$\gamma$ denotes the local vertical and is perpendicular to the plane formed by $\alpha$ and $\beta$
	(plane tangent to the local solar surface).
	Panels (a) and (b) show different perspectives of the filament trajectory (color-coded), where
	the black dashed-lines on the planes $\alpha$-$\beta$ depict the top-view trajectory of the
	filament eruption. The 3D projections reveal that the filament propagates markedly deviated from
	the local vertical direction. (An animation of this figure is available)
        }
        \label{fig:inc_3D}
\end{figure*}

The trajectory of the filament eruption (apex) calculated with Eq.~(\ref{Eq:matrix}) is presented in Fig.~\ref{fig:inc_3D}.
Panels (a) and (b) show different perspectives of the 3D space, which reveal that the filament propagates making a pronounced
angle with the local vertical ($\gamma$-axis) and travels more or less inclined towards the solar surface (see also the associated
animation). Panel (b) shows that during the initial process of the eruption (05:10--05:25~UT), the filament experiences small
variations in its propagation direction, thereafter, the filament gradually recovers its ascending motion but proceeds maintaining
a relatively large inclination angle with respect to the local vertical.
These findings demonstrate that the studied filament erupted following a non-radial direction and markedly deviated from the local
vertical direction. \cite{cabezas2017}, using multiwavelength imaging observation in the H$\alpha$ line, also reported on a
filament that erupted nearly horizontally exhibiting variations of the inclination angle.

\section{Discussion}\label{sect_discussion}

\subsection{Fast-filament eruption}
The filament eruption introduced in this paper occurred in a spotless region associated with a weak B-class solar flare. The
derived maximum true velocity $600~{\rm km~s^{-1}}$ and acceleration $2.5~{\rm km~s^{-2}}$ are considerably larger than the
commonly reported in previous works. This kind of observation represents a challenge for interpreting and modeling because the
common thought is that active region filaments are subject to experience fast acceleration.
Focusing on the Doppler velocity, \cite{li2005} reported on a complex filament eruption in the H$\alpha$ line blue-shifted over
$400~{\rm km~s^{-1}}$. Similarly, \cite{kleint2015} investigated a fast-filament eruption observed in transition region spectral
lines and obtained a maximum Doppler velocity of about $600~{\rm km~s^{-1}}$. The common aspect of the two mentioned fast
eruptions is that both originated from complex active regions and were associated with X-class flares.
According to the general picture, solar eruptions predominantly occur in magnetically complex active regions and the accumulated
magnetic energy associated with non-potential fields constitutes the primary source to power the eruptions. The studied filament
eruption, however, cannot adequately be explained with the above standard view.

If, however, we consider the unwinding characteristics of the filament eruption that was revealed by the observation
(see Fig.~\ref{fig:AIAtimeslice}, also the associated animation of Fig.~\ref{fig:mosaic}), we can hypothesize that the filament
was strongly twisted during the pre-eruption stage. Theoretical models of solar eruptions also invoke that the twisting mechanism
can be a source of energy injection \citep{sakurai1976}, and highly twisted flux ropes rise much faster releasing a large amount
of energy \citep[e.g.][]{amari2003, williams2005}. For example, \cite{nishida2013} by performing a three-dimensional MHD simulation
showed that strongly twisted flux rope experiences high ejection speed allowing a rapid acceleration.
On the other hand, \cite{kurokawa1987} examined a high-speed filament eruption in which unwinding and rotation characteristics
were observed in H$\alpha$ filtergrams and concluded that the eruption process can be explained with the magnetic twist model
proposed by \cite{shibata1986}. It has been also shown by \cite{moore1988} that the expansion and unwinding of the field lines in
and around the erupting filaments are directly related to the release of magnetic energy, which can supply enough power to drive
the eruptions. Therefore, we argue that the high-speed of the studied filament eruption can be a consequence of the initially
strong twist and the release of a considerable amount of accumulated energy (see Sect.~\ref{filament_mass}).
Here it is important to note that other mechanisms such as magnetic reconnection also could have contributed, because it was found
that mass acceleration and the reconnection rate are well correlated \citep[e.g.][]{qiu2004,lynch2008}.

\subsection{Doppler-shifted components and velocity distribution}\label{Doppler_components}  
The H$\alpha$ optical thickness profiles revealed that the erupting filament is made of multiple sub-components that are
manifested showing discrete peaks in the profiles (see Fig.~\ref{fig:ApexLatFastMain}). As the eruption develops, between 05:32
-- 05:37~UT the main peak (main bulk) of the fast (leading) and main (core) components at the filament apex (upper panels in
Fig.~\ref{fig:ApexLatFastMain}) are blue-shifted from about $-30$ to $-145~{\rm km~s^{-1}}$ and from $-36$ to
$-102~{\rm km~s^{-1}}$, respectively. Similarly at the lateral part of the erupting filament
(lower panels in Fig.~\ref{fig:ApexLatFastMain}), between 05:41 -- 05:44~UT, the main peak of the fast (leading) and main (core)
components are blue-shifted from about $-72$ to $-265~{\rm km~s^{-1}}$ and from $-70$ to $-184~{\rm km~s^{-1}}$, respectively.
The abrupt change in the velocity mentioned above happened exceptionally fast, within $\sim$5 minutes. In the meantime, some
sub-components that appear ahead of the main bulk are shifted even to much higher velocities, for example to nearly
$-400~{\rm km~s^{-1}}$ at 05:36:44 (apex fast) and 05:44:19~UT (lateral fast). The profiles of the sub-components are rather
narrow than the profiles of the main bulk, and their corresponding micro-turbulent velocities $\xi$ inferred from the model fitting
(see Sect.~\ref{sect_spectral_vel}) differ from each other.
We note that in some cases the fitting results yield large micro-turbulent velocities beyond those expected for quiescent filaments
and prominences \citep[e.g.][]{molowny1999, schwartz2019, okada2020, yamasaki2023}, even over the estimates of previous studies
for the case of eruptive events \cite[e.g.][]{labrosse2012, heinzel2016, zhang2019}.
One must note, however, that in the case of filament and prominence eruptions significant non-thermal motions should be expected
due to more turbulent plasma that could broaden the profiles. Regarding our observation, we argue that the pronounced turbulent
velocity could occur because of the filament fast expansion and the superposition of unresolved structures moving with distinct
velocities.
On the other hand, as it is also noted in Figs.~\ref{fig:Vlos_apex} and \ref{fig:Vlos_Lat} the line-of-sight velocity distribution
is greatly enhanced in the late phase of the eruption and the presence of sub-components becomes more conspicuous. \cite{wang2022}
also reported enhancements of the line-of-sight velocity distribution of an erupting filament, but only up to about
$-22~{\rm km~s^{-1}}$.

The characteristics described above indicate that complex plasma motions originate as a consequence of the fast eruption, leading
the individual blob-like structures to expand rapidly with random velocities. These findings also suggest that large velocity
dispersion is manifested during the main and final episodes of the eruption which could be related to the internal motions.
A likely explanation for the appearance of the sub-components or blobs moving with distinct velocities is that internal structures
existed in the filament body before the eruption, and the filament disruption allowed these confined structures to escape and
flow abruptly with large velocities. These characteristics are also well observed in the Doppler velocity maps
(see Fig.~\ref{fig:VLOS_maps} and the animation) where most of the sub-components or blob-like structures are ejected with
supersonic velocities. Although we cannot confirm in our data the existence of internal structures in the filament body before its
eruption, previous high-resolution observations showed evidence of complex internal structures right before the filament eruptions
\citep[e.g.][]{sasso2011, schwartz2019}, and also manifestations of internal motions in the spectral profiles during a filament
eruption \citep{penn2000}.

Another aspect that we would like to mention here concerns the fixed source function adopted to calculate the optical thickness of
the filament eruption in Sect.~\ref{sect_spectral_vel}. It has been shown by \cite{heinzel1987} that for prominence-like structures
located at greater heights above the solar surface and moving with large vertical velocities, the H$\alpha$ source function can   
substantially change due to the so-called Doppler brightening effect (DBE). In other words, there is a velocity effect on the     
H$\alpha$ line source function \citep[cf.][]{heinzel1999}.                                                                        
To evaluate quantitatively the influence of DBE on our fast-moving filament, we consider the height and velocity presented in
Fig.~\ref{fig:true_hva} and calculate the source function following \cite{heinzel2015} and \cite{heinzel2015b}. In the case of    
filaments seen in absorption against the background solar disk, the source function is mostly controlled by the photon scattering,
so one can approximate $S\approx W I_0$, where $W$ is the geometrical dilution factor that decreases with increasing height, and  
$I_0$ is the incident radiation expressed in terms of the central depression of the H$\alpha$ line $0.17$ multiplied by the       
disk-center continuum intensity $4.077\times10^{-5}~{\rm erg~s^{-1}~cm^{-2}~sr^{-1}~Hz^{-1}}$ \citep[see][]{david1961,heinzel1999}.
Focusing on the main phase of the eruption (see Fig.~\ref{fig:true_hva}), at 05:35~UT the filament is traveling with a velocity of
about $150~{\rm km~s^{-1}}$ located at a height of $5\times10^4~{\rm km}$. For this height, we find that the intensity of the     
incident radiation illuminating the moving cloud (filament) is diluted by a factor of $0.32$ \citep[see Eq.~(7) in][]{heinzel2015b}.
In addition, for the height and velocity mentioned previously, DBE approaches a factor of about $1.85$                            
\cite[see Fig. 11 in][]{heinzel1987}. Therefore, the source function influenced by DBE and normalized to the continuum level is   
$S=0.1$. This result is the same as our assumed value in Sect.~\ref{sect_spectral_vel}. We note that our choice of considering    
the results given in Fig.~11 of \cite{heinzel1987} is also because the computation was done for the case of a large micro-turbulent
velocity, such as we found in our observation.                                                                                    
To summarize, if DBE is neglected when calculating the source function, the result obtained above is reduced by a factor of two,  
and if large DBE factors (and thus large velocities) are included in the computation, the source function is moderately enhanced. 
However, as already mentioned in Sect.~\ref{sect_spectral_vel} our results of the optical thickness obtained with                 
Eq.~(\ref{Eq:tau2}) are not very sensitive to the variation of the source function, even though if the source function varies, the
determination of the Doppler-shifted plasma in the optical thickness spectra and thus the velocity, quantity that is our primary  
focus, is not affected.                                                                                                           

\subsection{Filament mass}\label{filament_mass}
Filament mass is an important physical quantity for considerations related to its formation, stability and dynamics. Here we
discuss on the probable total mass contained in the studied filament and provide a quantitative estimate. We take into account the
filament geometry and the line center optical thickness $\tau_0$ calculated with the fitting model in
Sect.~\ref{sect_spectral_vel}, as well as the non-LTE theoretical relations given in \cite{heinzel1994}. If we approximate the
filament body to a cylindrically symmetric flux tube, the filament length and the diameter of its corresponding cross-sectional
area at 05:37:47~UT (see first column of Fig.~\ref{fig:mosaic}) are $\sim$$1.8\times10^{10}~\rm{cm}$ and
$\sim$$1.3\times10^9~\rm{cm}$, respectively. At this time the optical thickness of the filament apex main component is
$\tau_0\simeq0.13$ (see Fig.~\ref{fig:ApexLatFastMain}).
On the other hand, the filament density can be calculated with $\rho_m = 1.4 m_H n_H$, where $m_H$ represents the mass of the
hydrogen atom and $n_H$ is the number density of the hydrogen including neutral atoms and protons \citep[cf.][]{heinzel2003}.
Using the correlation-plots given in Figs. 5, 8, and 23 of \cite{heinzel1994}, for $\log(\tau_0)\simeq0.13$ we get the line
integrated emission $\log(E_{H\alpha})\simeq5.0$, the electron density $\log(n_{e})\simeq11.0$, and then the total hydrogen
density $\log(n_H)\simeq10.9$, respectively. With the calculated $n_H$ the filament density is
$\rho_m\simeq2.12\times10^{-13}~{\rm g~cm^{-3}}$. Finally, considering the geometry and the dimensions given above the filament
mass is $\sim$$5.4\times10^{15}~{\rm g}$.
This result indicates that the studied filament is rather a relatively massive erupting structure. \cite{grechnev2014} investigated
a filament eruption in the H$\alpha$ line which exhibited similar morphological characteristics to our filament and obtained a
filament mass comparable to our result. 

Furthermore, taking into account the estimated filament mass we can also gain insights about the energy involved in the eruption.
Considering the scaling relations between mass ejection and flare energy given in Fig.~6 (b) of \cite{kotani2023}, we can argue
that the total flare energy associated with the filament eruption could be of the order of $\sim$$10^{31}{\rm erg}$. If we compare
our estimated filament mass and the associated flare energy with the results shown in Fig.~3 of \cite{namekata2021} for the case
of solar filament eruptions, we can see that our values lie above the results of previous studies. In Paper II, we provide further
discussions about the filament total mass and the associated energetics.

\subsection{Downflows driven by the filament eruption}
We have identified intermittent downflows as red-shifted plasma close to the footpoints of the erupting filament whose
corresponding velocities range from $45$ to $125~{\rm km~s^{-1}}$ (see lower panels of Fig~\ref{fig:mosaic}). Our interpretation
is that the observed red-shifted signatures are caused by plasma flows along the filament legs that are precipitating down to the
solar surface. The downward-moving plasma represents drainage of the filament material, which evidently has direct consequences on
the dynamics of the filament.
The downflow velocity we found is larger than in previous studies. For example, \cite{doyle2019} using H$\alpha\pm1.38~{\AA}$
observations found downflows along the legs of an erupting filament red-shifted up to $45~{\rm km~s^{-1}}$.
\cite{guo2023} also detected downward plasma flows during a filament eruption red-shifted to about $20~{\rm km~s^{-1}}$. These
authors concluded that the red-shifted plasma was due to mass draining towards the solar surface.
The large velocity we detected in our data is simply because the spectral coverage of our instrument is sufficient enough to
capture high-speed structures associated with the filament eruption. Relatively large velocity of mass drainage in erupting
filaments was also reported. \cite{dai2021} found velocities in the range of $35-85~{\rm km~s^{-1}}$ of draining material,
although these measurements were obtained based on the projected motion in the plane-of-sky.
We conclude that the observed downflows that appear during the filament eruption constitute mass loss of the filament that might
have contributed to the triggering of the fast acceleration. A model of an expanding magnetic loop that incorporates the temporal
variation of the H$\alpha$ spectrum was developed by \cite{ikuta2024} recently. This model can be suitable to describe the
appearance of downflows in erupting filaments, such as we found in our observation.
Alternatively, the red-shifted plasma flows detected close to the footpoints of the filament eruption could be further
enhanced due to the expansion and helical motions of the erupting material \citep{otsu2024}.

\subsection{Implication for models}\label{implication-model}
One of the important results of this work was the detection of highly Doppler-shifted plasma that accompanied the filament
eruption. The large blue-shifted plasma (see Figs.~\ref{fig:VLOS_maps}, \ref{fig:Vlos_apex}) indicates that the filament material
was launched into the outer space having a significant line-of-sight component.
This result is a clear example that the actual velocity and propagation direction of filament eruptions could be greatly
influenced by the line-of-sight component and the fact of not including it in the computations would certainly yield inaccurate
results. This is also an important issue for modeling solar eruptions. Since in our current understanding erupting filaments and
prominences form part of the flux rope embedded in the CMEs core \cite[e.g.][and references therein]{shibata2011, webb2015,
heinzel2016, wood2016, chen2017, green2018}, models attempting to predict the early propagation of CMEs will have to include as
initial boundary conditions accurate estimates taking into account the contributions of all velocity components
(e.g. tree-dimensional velocity field).
Another important point to mention here is that filaments and prominences do not always erupt in radial direction or perpendicular
to the solar surface, such as assumed in many models. This is a subject of great importance that is not often discussed. As shown
in Fig.~\ref{fig:inc_3D} \cite[see also Fig.~6 in][]{cabezas2017}, filaments can significantly deviate from the radial direction.
In the case of the studied filament, we speculate that the erupting material possibly experienced a change in its rise direction
when interacting with the overlying asymmetric magnetic field and its surroundings. 
Other case studies showing that the trajectory of erupting prominences is generally not in the radial direction were presented by
\cite{zapior2010} and \cite{zapior2016}. Indeed, the change in the propagation direction of filament and prominence eruptions may
also have implications for the initial trajectory of CMEs \cite[e.g.][]{lugaz2011, kliem2012, guo2023}, which we do not address in
this article.

On the other hand, the derived maximum true velocity and acceleration (Fig. \ref{fig:true_hva}) may have been even much larger.
As shown in Fig. \ref{fig:AIAtimeslice} (b), in the He~{\sc ii} line the filament continues ascending to greater distances even
after it has disappeared in H$\alpha$. However, we cannot confirm the actual velocity of the eruption beyond 05:40~UT because of
the lack of spectroscopy observations in spectral lines other than H$\alpha$.
Looking at the time-distance profile (yellow plot) we suggest that during the final stage of the eruption the filament apparently
underwent a drastic diminution in the velocity, meaning that the main acceleration process occurred earlier than 05:40~UT, that is,
the period at which the event is well observed in H$\alpha$. Hence, the true eruption profiles obtained solely based on H$\alpha$
data are adequate to explain most characteristics of the studied filament eruption.
Nevertheless, it would be advantageous to have simultaneous full-disk imaging spectroscopy observations in various spectral lines,
for instance, H$\alpha$ and He~{\sc i} (10830~{\AA}), which could provide a better picture of erupting filaments at different
heights of the solar atmosphere, also it would be beneficial for modeling.

Last, numerical modeling will certainly be important to deepen our understanding of the physics behind the fast eruptions. For
example, how much energy is necessary to power the studied fast-filament eruption? What about the action of the driving forces?
And how does the mass loss influence the acceleration process of the filament?
In Paper~II of this series we address these questions and provide quantitative estimates.

\section{Conclusions}

In this paper, we have investigated a fast-filament eruption that occurred on 2017 April 23. We used H$\alpha$ imaging
spectroscopy observations provided by SMART/SDDI, which allowed us to characterize the whole eruption process in detail. Results
of the ``cloud model'' fitting and spectral diagnostics revealed that the filament was ejected with a line-of-sight velocity
exceeding $-250~{\rm km~s^{-1}}$. This is an important finding and demonstrates that high-resolution full-disk imaging
spectroscopy observations are essential to detect high-speed mass motions ejected from the Sun that potentially could impact the
space environment at Earth. Most existing solar instruments generally miss large line-of-sight velocities of filament and
prominence eruptions, partly due to the limited spectral coverage or small field of view.

The spectral analysis showed that the erupting filament is composed of multiple sub-components moving with discrete velocities.
This suggests that internal complex structures existed before the filament eruption. Detailed sub-arcsecond resolution
observations are necessary to reach conclusions on the relation between the sub-components and the detected large velocity. On the
other hand, the distribution of the optical thickness profiles allowed us to apply a different approach to calculate the
line-of-sight velocity profiles, enabling us to derive the true velocity and acceleration profiles of the filament eruption. This
alternative method may be suitable for obtaining time velocity profiles  of filament and prominence eruptions captured in
spectroscopic observations. Moreover, thanks to the velocity vectors it was also possible to reveal that the filament erupted
making a pronounced angle with the local vertical direction.

Finally, in the context of space weather, it is crucial the continuous monitoring of potential source regions on the solar disk and
build procedures to accurately infer their physical properties (e.g. absolute velocity and propagation direction). The collected
and inferred information can eventually be used as initial conditions to predict, for example, the initial propagation of solar
eruption and Earth-directed CMEs.
In line with the above statements, the Astronomical Observatory of Kyoto University is leading and promoting the Continuous
H-Alpha Imaging Network \citep[CHAIN;][]{ueno2007} project aiming to obtain continuously synoptic Doppler maps and to provide the
three-dimensional velocity field of filament/prominence eruptions and associated events
\citep[e.g.][]{cabezas2017,cabezas2019,seki2019,gutierrez2021}. Currently, three instruments of the CHAIN project are in operation
at three strategic locations around the globe, SMART/SDDI at Hida Observatory (Japan), and two Flare Monitoring Telescopes, one
located at San Luis Gonzaga National University (Ica, Peru) and the other at King Saud University (Saudi Arabia). It is also important
to mention that several efforts are going in the same direction, such as the next-generation GONG network \citep{pevtsov2022}, the
project of Synoptic Solar Observations at NAOJ \citep{hanaoka2020}, and the recently launched space
mission Chinese H-alpha solar explorer \citep[CHASE;][]{li2022}.

\begin{acknowledgements}
	The authors are grateful to the staff members of Hida Observatory for the operation and continuous solar observations
	with the SMART/SDDI. D.P.C. is thankful to Shin-ichi Nagata, Yuwei Huang, Takako T. Ishii, and Kumi Hirose for fruitful
	discussions, He also thanks KD Leka for insightful comments and discussions. D.P.C gratefully acknowledges support from
	Kyoto University during the post-doctoral fellowship at Hida Observatory. This work is supported by the Institute for
	Space-Earth Environmental Research (ISEE), Nagoya University.
	This work was also supported by JSPS KAKENHI grant Nos. JP24K07093 (PI: A. A.) JP21H01131 (PI: K. S.), and JP24K00680
	(PI: K. S.).
	AIA data products are courtesy of NASA/SDO science team, a mission for NASA's Living With a Star program. In this
	research we have made use of Python packages SunPy, Astropy, Scipy, Matplotlib, and Numpy. Additionally, we used the
	{\it Solarsoft} library.
	Finally, we appreciate the constructive comments and suggestions of the anonymous referee that helped improve the clarity
	and presentation of this work.
\end{acknowledgements}

%
%


\begin{appendix} 
\section{Distribution of the optical thickness at the lateral part of the filament eruption}

\begin{figure}[ht]
\includegraphics[width=10.0cm]{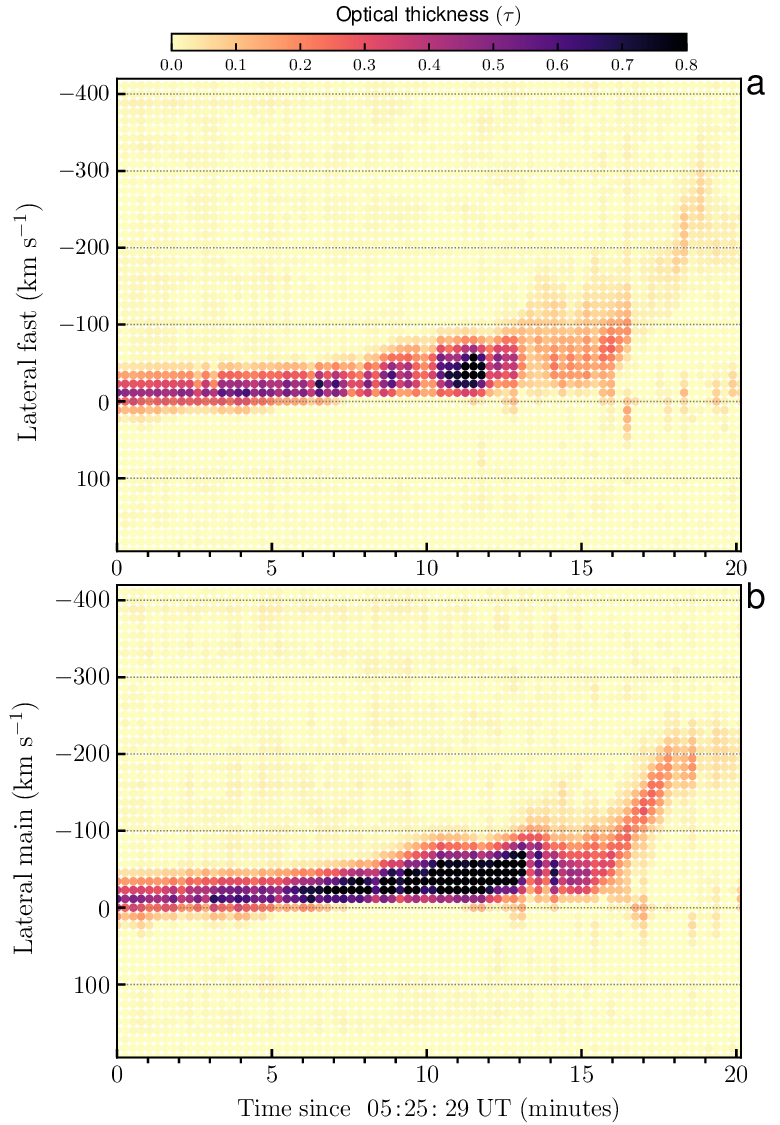}
	\caption{Same as in Fig.~\ref{fig:Vlos_apex} but for the lateral part of the filament eruption
	(main and fast components). The distribution plots in panel (a) clearly show that the erupting
	material is blue-shifted to velocities of about $-300~{\rm km~s^{-1}}$.}
\label{fig:Vlos_Lat}
\end{figure}
\end{appendix}

\end{document}